\documentclass[9pt]{acmart}
\usepackage[colorinlistoftodos]{todonotes}
\usepackage[ruled,vlined,linesnumbered]{algorithm2e}
\usepackage{subfigure}
\SetKwInput{kwIn}{Input}
\SetKwInput{KwOut}{Output}
\SetKwInput{KwInit}{Initialization}
\usepackage{algorithmic}
\newtheorem{newexample}{Example}[section]
\usepackage{mathtools}
\DeclarePairedDelimiter\ceil{\lceil}{\rceil}

\newcommand\revise[1]{%
  \protect\leavevmode
  \begingroup
    \color{black}%
    #1%
  \endgroup
}
\newcommand{\dbo}{RDBMS-X
}
\newcommand{\dbm}{RDBMS-Y
}

\newcommand{\dboSmall}{rdbmsX}
\newcommand{\dbmd}{rdbmsY}
\newcommand{\dboSmallIM}{rdbmsX\_im}
\newcommand{\dbmnon}{rdbmsY\_non}

\AtBeginDocument{%
  \providecommand\BibTeX{{%
    \normalfont B\kern-0.5em{\scshape i\kern-0.25em b}\kern-0.8em\TeX}}}

\copyrightyear{2021}
\acmYear{2021}
\setcopyright{rightsretained}
\acmConference[SIGMOD '21]{Proceedings of the 2021 International
Conference on Management of Data}{June 20--25, 2021}{Virtual Event,
China}
\acmBooktitle{Proceedings of the 2021 International Conference on
Management of Data (SIGMOD '21), June 20--25, 2021, Virtual Event,
China}
\acmDOI{10.1145/3448016.3457314}
\acmISBN{978-1-4503-8343-1/21/06}

\begin{document}
\fancyhead{}
\title{Vertex-centric Parallel Computation of SQL Queries \\
Extended Version}


\author{Ainur Smagulova}
\affiliation{
  \institution{UC San Diego}}
\email{asmagulo@eng.ucsd.edu}

\author{Alin Deutsch}
\affiliation{\institution{UC San Diego}}
\email{deutsch@eng.ucsd.edu}



\begin{abstract}
We present a scheme for parallel execution of SQL queries
on top of any vertex-centric BSP graph processing engine.
The scheme comprises a graph encoding of relational instances
and a vertex program specification of our algorithm called TAG-join,
which matches the theoretical communication and computation complexity of
state-of-the-art join algorithms. 
\revise{When run on top of the vertex-centric TigerGraph database engine on a single multi-core server,
TAG-join exploits thread parallelism and is competitive with (and often outperforms)
reference RDBMSs on the TPC benchmarks they are traditionally
tuned for. In a distributed cluster, TAG-join outperforms the popular Spark SQL engine.}
\vspace{-2mm}
\end{abstract}

\settopmatter{printfolios=true}
\maketitle

\section{Introduction}

We study the evaluation of SQL join queries
in a parallel model of computation that we show to be extremely well-suited
for this task despite the fact that it was designed for a different purpose
and it has not been previously employed in this setting.
We are referring to the vertex-centric flavor~\cite{Malewicz10}
of Valiant's bulk-synchronous parallel (BSP) model of computation~\cite{Valiant90}, 
originally designed for processing analytic tasks over data modeled as a graph.

Our solution comprises (i) a graph encoding of relational instances which we call
the {\em Tuple-Attribute Graph (TAG)}, and (ii) an evaluation algorithm specified as a vertex-centric program running over TAG inputs. 
The evaluation is centered around a novel join algorithm
we call {\em TAG-join}.

On the theoretical front, we show that TAG-join's
communication and computation complexities are competitive
with those of the best-known parallel join algorithms~\cite{afrati_1,koutris_1,
koutris_2,hu_1,hu_2,koutris_book} while avoiding 
the relation reshuffling these algorithms require (for re-sorting or re-hashing)
between individual join operations.
TAG-join adapts techniques from the best sequential
join algorithms (based on worst-case optimal bounds 
\cite{nprr,leapfrog,skew_strikes}
and on generalized hypertree decompositions~\cite{gottlob_1,gottlob_2}), 
matching their computation complexity as well.

\revise{
On the practical front, we note that our vertex-centric SQL evaluation scheme applies to both intra-server thread parallelism and to distributed cluster parallelism. 
The focus in this work is to tune and evaluate how our approach exploits thread parallelism 
}
in the \textit{"comfort zone"} of RDBMSs: 
running the benchmarks they are traditionally tuned for,
on a multi-threaded server with large RAM and SSD memory holding all 
working set data in warm runs.

We note that
the benefit of recent developments in both parallel and
sequential join technology has only been
shown in settings beyond the RDBMS comfort zone. 
The parallel join algorithms target scenarios of clusters with numerous processors, 
while engines based on worst-case optimal algorithms tend to be outperformed 
\footnote{Their benefit kicks in on queries where intermediate results are much larger than 
\revise{the input tables. This is not the case with the primary-foreign key joins that are prevalent in OLTP and OLAP workloads since the cardinality of $R \bowtie_{R.FK = S.PK} S$ is upper bounded by that of $R$ (every $R$-tuple joins with at most one $S$-tuple)}.}
by commercial RDBMSs operating in their comfort zone~\cite{level_headed,eh,subgraph_wcoj,newmann}. 

TAG-join proves particularly well suited
to data warehousing scenarios (snowflake schemas, primary-foreign key joins).
Our experiments show competitive performance
on the TPC-H~\cite{tpch} and across-the-board dominance on the TPC-DS~\cite{tpcds} benchmark.

\revise{
In a secondary investigation, we also evaluate our TAG-join implementation's
ability to exploit parallelism in a distributed cluster, showing that
it outperforms the popular Spark SQL engine~\cite{spark_sql}.
}

A bonus of our approach is its \revise{applicability on top of} 
vertex-centric platforms without having to change their internals.
There are many exemplars
in circulation, including open-source \cite{giraph, graphlab,powergraph,graphx}
and commercial~\cite{Malewicz10,TG19}.
We chose the free version of the TigerGraph engine~\cite{TG19,tg} for our evaluation
due to its high performance.

Our work uncovers a synergistic coupling between the TAG representation of relational databases and vertex-centric parallelism that went undiscovered so far because,
despite abundant prior work on querying graphs on native
relational backends~\cite{case_against, all_in_one,vertexica,thewheel}, 
there were no attempts to query relations on native graph backends.

\paragraph{Paper organization}
After reviewing the vertex-centric BSP model in Section~\ref{sec:vc-model},
we present the TAG encoding of relational instances (Section~\ref{sec:tag}), then develop TAG-join starting from two-way (Section~\ref{sec:vc-two-way}), to acyclic (Section~\ref{sec:acyclic_multiway}) and to arbitrary join (Section~\ref{sec:arbitrary}). We 
discuss extensions beyond joins in Section~\ref{sec:beyond-joins}, report experiments in Section~\ref{sec:experiments} and conclude in Section~\ref{sec:conclusion}.

\section{Vertex-centric BSP Model}
\label{sec:vc-model}
The Vertex-centric computational model was introduced by 
Google's Pregel~\cite{Malewicz10} system as an adaptation to graph data of 
Valiant's Bulk Synchronous Parallel (BSP) model of computation \cite{Valiant90}. 

A BSP model includes three main components: a number of processors, 
each with its own local memory and ability to perform  local computation; 
a communication environment that delivers messages from one processor to another; 
and a barrier synchronization mechanism. 
A BSP computation is a sequence of supersteps. 
A superstep comprises of a computation stage, where each processor performs a sequence of operations on local data, and a communication stage, where each processor sends a number of messages. The processors are synchronized between supersteps, i.e. they
wait at the barrier until all processors have received their messages.

The vertex-centric model adapts the BSP model to graphs, 
such that each vertex plays the role of a processor that executes a user-defined program. Vertices communicate with each other by sending messages via outgoing edges, or directly to any other vertex whose identifier they know (e.g. discovered during computation).

Each vertex is identified by a vertex \revise{ID}. It holds a state,
which represents intermediate results of the computation; a list of outgoing edges; 
and an incoming message queue. Edges are identified by the ids of their source and destination vertices, and they can also store state. The vertex program is designed from the perspective of a vertex. The vertex program operates on local data only: the vertex state, 
the received messages, and the incident edges. 

At the beginning of a computation all vertices are in active state, and start the computation. 
At the end of the superstep each vertex deactivates itself, and it will stay inactive unless it receives messages. All messages sent during superstep $i$ are available at the beginning of superstep $i+1$. Vertices that did not receive any messages are not activated in superstep $i+1$, and thus do not participate in the computation. 
The computation terminates when there are no active vertices, i.e. no messages were sent during the previous superstep. The output of the computation is the union of values computed by multiple vertices (distributed output). 

\paragraph{Aggregators}
Aggregators provide a mechanism for vertices to collaborate in order to compute a global aggregate value.
This mechanism is defined as an aggregation vertex, whose id is known to all the vertices in the graph and they can send messages directly to it.
The aggregation vertex then aggregates the received values, and can share the computed value by sending messages back to the vertices, e.g. to be used further as a condition to trigger the next computation phase, or deactivate (eliminate) a vertex. There can be multiple aggregation vertices defined for the computation. 

\paragraph{Cost Measure}
We measure the total communication and computation cost of an algorithm. The \textit{total communication cost} is the sum of all the messages sent by vertices over all supersteps. We do not include the received messages in to the cost, since any outgoing message is the incoming message of at least one vertex, and thus it's sufficient to count it once. The \textit{total computation cost} is the sum of the amount of computation performed by vertices over all supersteps. We account for computation cost to make sure that when designing algorithms each vertex performs limited work and does not exceed the communication cost.

\paragraph{Examples of vertex-centric engines} A vertex-centric BSP model was first introduced in Pregel \cite{Malewicz10}, 
\revise{followed by a proliferation of
open-source and commercial implementations, some running on distributed clusters,
others realizing vertex communication  via a shared memory.
Surveys of the landscape can be found in~\cite{mccune15,yan17}, while
their comparative experimental evaluation has been reported in~\cite{ammar18,han14,lu14}.
These works exclude the new arrival TigerGraph~\cite{TG19}, which exploits both
thread parallelism within a server and distributed cluster parallelism.
}

\section{TAG Encoding of a Relational DB}\label{sec:tag}
\begin{figure*}[]
  \centering
  \includegraphics[width=\linewidth]{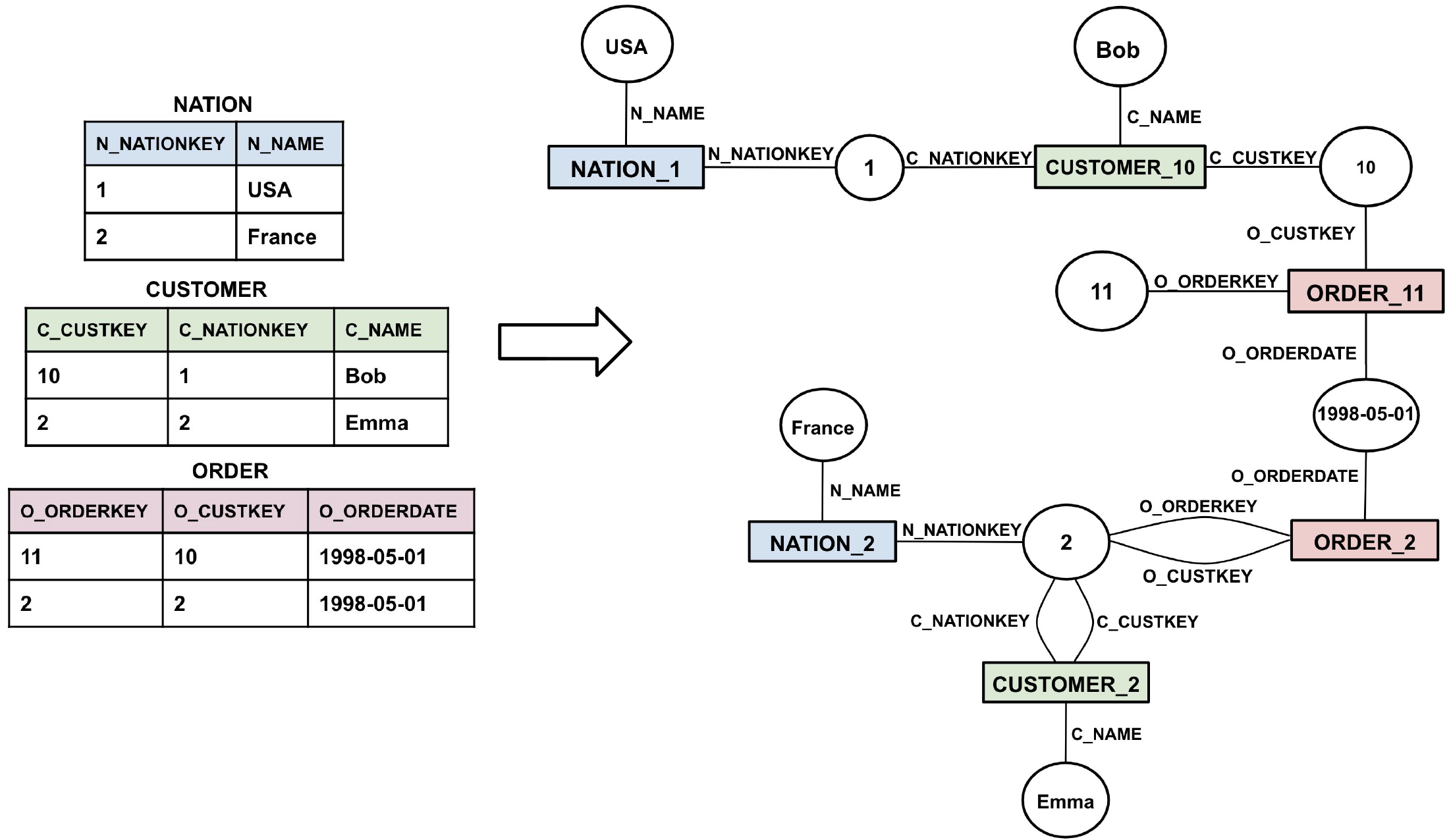}
  \caption{Encoding relational data in a TAG representation. Tuple vertices are depicted as rectangles, and attribute vertices as circles.}
  \label{data_model}
  \Description{}
\end{figure*}
We present the Tuple-Attribute Graph (TAG) data model we use to encode a relational database as a graph.
A graph is a collection of vertices and edges, where each vertex and edge has a label and can store data as a collection of (key,value) pairs (i.e. attributes).
The TAG model defines two classes of vertices: {\em tuple} vertices, representing tuples of a relation; and {\em attribute} vertices, representing attribute values of a tuple.
Tuple and attribute vertices function the same in the vertex-centric computational model, i.e. both can execute a user-defined program and communicate via messages.

We construct a TAG graph from a relational database as follows.
\begin{enumerate}
\item For every tuple $t$ in relation $R$ 
create a tuple vertex $v_t$ labeled $R$ (each duplicate occurrence
of $t$ receives its own fresh tuple vertex). Store $t$ 
in $v_t$'s state. 
\item For each attribute value $a$ in the active domain of the 
database create an attribute vertex $v_a$. 
Add a label based on the domain/type of $a$ (e.g. int, string, etc.). 
Create exactly one vertex per value regardless
of how many times the value occurs in the database.
\item For each occurrence of $R$-tuple $t$ with an attribute named $A$ of value $a$,
add an edge labeled $R.A$ between $v_t$ and $v_a$.
\end{enumerate}
Notice that the graph is bipartite, as edges never connect tuple vertices with each other,
nor attribute vertices with each other.

\begin{newexample}
Figure \ref{data_model} shows the example instance of relational data and its corresponding graph representation (to unclutter the figure, edge labels do not include the table names). We start with the first tuple of relation $NATION$, and map it to a tuple vertex with id $NATION\_1$ with corresponding label $NATION$. Then each of its attribute values maps to an attribute vertex, i.e. value $1$ maps to an integer attribute vertex, and value $USA$ maps to a string attribute vertex. We finish the transformation by creating edges from the tuple vertex to the two attribute vertices with labels that correspond to the attribute names of the tuple. We repeat the same steps for the rest of the tuples and for all relations. Tuple vertex $CUSTOMER\_10$ also uses integer value $1$ as its attribute, and thus we simply add an edge to connect it to this attribute vertex. Note how integer attribute vertex $2$ is shared among three tuple vertices $NATION\_2, CUSTOMER\_2, ORDER\_2$, and used as a value of five different attributes, hence the five edges with different labels are added. Two $ORDER$ tuples are connected with each other via the date attribute value that they share.
\hfill$\Box$
\end{newexample}
\noindent
The TAG representation of a relational database is query-independent
and therefore can be computed offline. Moreover, the size of the graph is linear in the size of the relational database. 

The most prominent feature of the TAG representation is that pairs of joining\footnote{\revise{In this paper, the
unqualified term "join" is shorthand for "equi-join".}} tuples are explicitly connected with each other via edges to their join attribute value.
For each attribute value $a$, 
the tuples that join through $a$ can be found by simply following the outgoing edges from the attribute vertex representing $a$.
Therefore, attribute vertices act as an indexing scheme for speeding up
joins. This scheme features
significant benefits over RDBMS indexing:

First, note that the TAG representation corresponds in the relational setting
to indexing {\em all} attributes in the schema. While this is prohibitively expensive 
in an RDBMS because of the duplication of information across indexes,
notice that TAG attribute vertices are not duplicated.
One can think of them as shared across indexes, in the sense that even if 
a value appears in an $A$ and a $B$ attribute, it is still represented only once
(e.g. attribute vertex 2 in Figure \ref{data_model}).

Second, an attribute vertex can lookup the tuples joining through it in
time linear in their number by simply following the appropriate edges.
In contrast, even when an appropriate RDBMS index exists (which cannot be taken for granted), 
the RDBMS index lookup time depends, albeit only logarithmically, on the size of the involved
input relations even if the lookup result size is small.

Third, attribute vertices are cheaper to build, 
and in the presence of changing data they are less challenging to maintain than
traditional RDBMS indexes, as they do not require any reorganization of the graph. 
It suffices to locally insert/delete vertex attributes and their incident edges.

Finally, since the set of edges is disjointly partitioned by the attribute vertices they 
are incident on, the TAG model is particularly conducive to parallel join processing
in which attribute vertices perform the tuple lookup in parallel. 
The vertex-centric BSP model of computation suggests itself as a natural candidate
because it enables precisely such parallel computation across vertices and messaging
along edges. 

In the remainder of the paper, we exploit this fact by developing a vertex-centric BSP join algorithm.~\footnote{
We represent TAG edges as undirected to merely indicate that each edge is a two-way relationship, and thus messages can be sent in both directions. 
To support messaging across directed out-edges in a standard vertex-centric program,
each undirected edge is modeled as two directed edges.}

Although we do not create duplicate attribute vertices for the same value, but let tuple vertices share attribute values that may even correspond to different attribute names in each tuple (e.g. note attribute vertex 2 in Figure \ref{data_model}), materializing attribute values still comes at cost of a bigger storage space. This can be mitigated in practice, by avoiding materializing some types of attribute vertices. For example, (a) when attribute is not likely to be used as a join condition, e.g. text value; or (b) when attribute values belong to a domain that is tricky to compare with equality operator, e.g. floats. We can store these values as attributes of a corresponding tuple vertex. A possible approach to more efficiently materialize float values in order to account for different precision and scale is to use a vertex per range of float values instead of a vertex per value. We do not consider the subject of mapping 'tricky' domain values to vertices in the scope of this paper, but simply avoid materializing these types of attributes in our experiments. It is important to note, in the interest of fairness, that if we were to create indexes on all attributes of all relations in any RDBMS, we would run into the same issues in terms of storage space, equality of floating-point numbers, and then index maintenance cost in addition.

\subsubsection{Related Encodings for Attribute-centric Indexing}
\revise{Paper~\cite{stoica19} addresses mapping of a relational instance to a property graph,
connecting vertices based on key-foreign key relationships only.
In contrast, the TAG encoding supports arbitrary equi-join conditions.
Moreover, \cite{stoica19} focuses on the data modeling aspect only
and does not address query evaluation (let alone the vertex-centric kind).

TAG encoding's attribute vertices generalize the value-driven indexing of~\cite{fletcher09} 
from RDF triples to arbitrary tuples. Their indexing role 
is also related in spirit to indexing Nested Relational data~\cite{deshpande88}.
Both works propose secondary indexing structures, with the requisite space and time
overhead for creation and maintenance. These are avoided in
TAG encoding, where attribute vertices are not redundant indexes, but the original data. Moreover, neither of~\cite{fletcher09,deshpande88} considers parallel evaluation 
of joins in general, and in particular the vertex-centric computational model.
}

\begin{figure*}[]
  \centering
  \includegraphics[width=\linewidth]{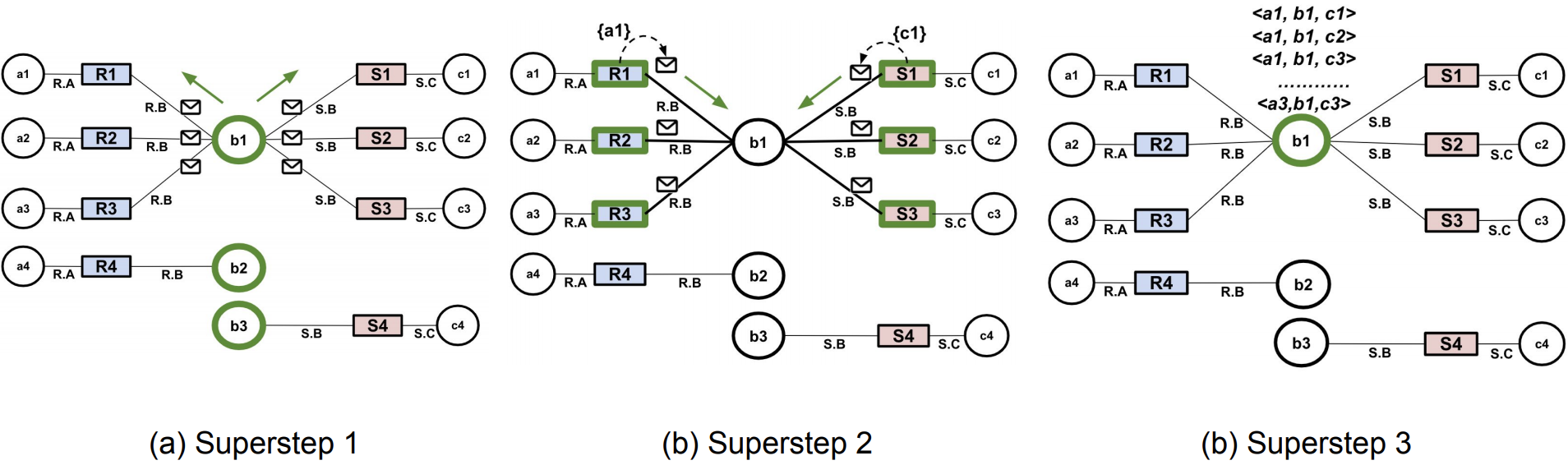}
  \caption{Example of a two-way join algorithm of relations R and S. Borders of 
  active vertices are highlighted in lighter shade (green).}
  \label{two_join}
  \Description{}
\end{figure*}

\section{Vertex-centric Two-Way Join}\label{sec:vc-two-way}

We begin with computing a natural join between two binary relations,
\begin{displaymath}
R(A,B) \Join S(B,C),
\end{displaymath}
using a vertex-centric algorithm over their TAG representation.
For presentation simplicity we consider natural join queries as examples throughout the 
paper, however note that any equi-join query can be transformed into a natural join by appropriate renaming of attributes. Furthermore, the generalization to n-ary relations
is straightforward.

Our approach is based on Yannakakis' algorithm for acyclic queries \cite{yannakakis}, that first performs a semi-join reduction on all relations to eliminate dangling tuples, and then joins the relations in arbitrary order. The semi-join is an essential query processing technique that can reduce the cost of a join query evaluation, especially in a distributed environment \cite{semijoin}. 

The semi-join of $R(A,B)$ with $S(B,C)$, denoted
$R \lJoin S$, retrieves precisely those 
$R$-tuples that join with at least one $S$-tuple.
To implement a reduction of $R$, it suffices to obtain a duplicate-free projection of 
the join column $S.B$: $R \lJoin S = R \lJoin \pi_{B}(S)$. 

Per Yannakakis's algorithm, in order to compute a two-way join we first reduce the sizes of $R$ and $S$ via a series of semi-joins, $J_1 := R \lJoin S$ and 
$J_2 := S \lJoin R$.  Now that the tuples that do not contribute to the output (a.k.a. 'dangling tuples') are removed, the algorithm constructs the join of the reduced relations, $J_1 \Join J_2$. 

Our algorithm follows a similar idea of splitting the computation into two phases: (1) a 
{\em reduction} phase to eliminate vertices and edges such that the surviving ones
correspond to the TAG representation of the reduced relations; and (2) a
{\em collection} phase that traverses the reduced TAG subgraph to collect vertex values
and construct the final join output. 

\subsection{Join on a Single Attribute}\label{sec:single_attrib_join}
By design of the TAG model, the semi-join reduction can be naturally mapped to a vertex-centric computational model. Each value $b$ of an attribute $B$ is mapped to a TAG 
attribute vertex $v_b$ whose outgoing edges connect it to precisely the (vertex representations of) $R$- and $S$-tuples that join through $v_b$. 
Thus, for a reduction to be performed, each attribute vertex needs to check its outgoing edges, confirming that it connects to at least one tuple vertex of each label $R$ and $S$, 
and then signal those tuple vertices that they are part of the final output by sending a message. Note that all attribute vertices can do this in parallel, independently of each other, since their individual computation and messaging only requires access to 
the locally stored vertex data. Also note that the attribute vertex need not 
'cross the edge' to inspect the proper labeling of the tuple vertex at the other end:
this information can be encoded in the edge label itself, by qualifying the attribute
name with the relation name it belongs to (see Figure~\ref{two_join}).

\subsubsection{Algorithm}\label{sec:two_way_algo}
We sketch a vertex-centric two-way join algorithm through the following example. 
Consider the TAG instance depicted in Figure \ref{two_join}. The computation starts by activating attribute vertices corresponding to the join attribute $B$.\\

\textbf{Superstep 1:} Each attribute vertex checks whether it can serve as 
\revise{\textit{join value}, i.e. a value in the intersection of columns R.B and S.B.}
For a vertex $v$ to be a join value it needs to have outgoing edges with labels R.B and S.B.
If this is the case, $v$ sends messages to the target tuple vertices via those edges. 
Otherwise, $v$ just deactivates itself. 
As depicted on Figure \ref{two_join}(a), vertex $b1$ figures out that it is a join value, and sends its id to the target tuple vertices (3 R vertices and 3 S vertices). Vertices
$b_2$ and $b_3$ deactivate themselves without sending any message.

\textbf{Superstep 2:} Tuple vertices are activated by the incoming messages. 
Each tuple vertex 
$t$ marks the edges along which it has received messages.
Then $t$ sends its non-join attribute value back to the join attribute vertex via the marked edges, and deactivates itself. In Figure \ref{two_join}(b), the $R$ tuple vertices 
$R1,R2,R3$ send their $A$ values and the $S$ tuple vertices $S1,S2,S3$ send their
$C$ values to vertex $b1$ via marked edges (shown in bold).

\textbf{Superstep 3:} Each active attribute vertex $v$
constructs a join result by combining the values received from both sides with their own
value (the operation is really a Cartesian product). Next, $v$
stores this result locally (or possibly outputs it to the client application). 
The computation completes when all vertices are deactivated. See Figure \ref{two_join}(c),
which shows the locally stored join result.\\

Superstep (1) corresponds to a reduction phase that eliminates tuple
vertices that do not contribute to the join. Supersteps (2) and (3) are part of a collection phase whose purpose is to collect, via messages, data from $R$-tuples and $S$-tuples 
that join and construct the output tuples.

In Superstep (3),
the $A$-attribute and $C$-attribute values received at a  $B$-attribute vertex
correspond to the {\em factorized representation} of the join result~\cite{fdb_2}, i.e.
the latter can be obtained losslessly as their Cartesian product. If the
distributed and factorized representation of the join is required, then
Superstep (3) is skipped.

\subsubsection{Cost Analysis}
Let $|R|$ and $|S|$ denote the sizes (in tuple count) of R and S respectively.
Then the total input size $IN = |R| + |S|$. Let $OUT = | R \Join S|$ 
denote the size of the join result
in the standard unfactorized bag-of-tuples representation.

In superstep (1), all $B$-attribute vertices are active (their number is upper bounded
by $IN$).
They send messages along the edges with labels R.B and S.B, but only when
the recipient tuples contribute to the join. 
Since the value of the join attribute disjointly partitions 
$R$, $S$, and also $R \Join S$, the total message count over all attribute
vertices is $|R \lJoin S| + |S \lJoin R|$, which is upper bounded by both
$IN$ and $OUT$, and therefore by $min(IN,OUT)$.
Even for the worst-case instance, where an attribute vertex is connected to all $R$ and $S$ tuple vertices, the communication cost does not exceed $min(IN,OUT)$. 
During the computation phase each attribute vertex with label $B$ (e.g. $b1,b2,b3$) iterates over its outgoing edges in order to figure out whether it joins tuples from both
$R$ and $S$. The total computation cost 
summed up over all vertices is upper bounded by the total number of edges and therefore 
by $IN$. 

In Superstep (2), only tuple vertices that contribute to the final output send messages,
for a total number of messages
$|R \lJoin S|$ (sent by $R$-labeled tuple vertices) 
+ 
$|S \lJoin R|$ (sent by $S$-labeled tuple vertices). 
This is again upper bounded by $min(IN,OUT)$.
Since the computation at tuple vertices is constant-time,
the overall computation in this superstep is also 
upper bounded by $min(IN,OUT)$.

In Superstep (3) each join vertex combines the received messages to construct output tuples. Each attribute vertex of value $b$ receives the $A$ attribute values
from its R-tuple neighbors, the $C$ attribute values from its $S$-tuple neighbors,
and computes $\sigma_{B=b}(R \Join S)$ locally as a Cartesian product.
The total computation cost across all active vertices is $OUT$
because the output tuple sets are disjoint across join vertices.
If the output is left distributed over the vertices
(the standard convention in distributed join algorithms~\cite{koutris_1,afrati_1}
is to leave the result distributed over processors, which in our setting are the vertices) 
no further messages are sent and the communication cost is 0. 
Even if the output is instead sent to a client application, 
the additional communication cost totalled over all vertices is $OUT$.
Observe that each join attribute vertex receives via messages
the {\em factorized representation} of the join output. 
This can potentially be significantly smaller than the size $OUT$ of the standard bag-of -tuples representation (in the worst-case instance, the factorized representation of the join result is $|R| + |S|$ while the unfactorized one is $|R| \times |S|$)
\footnote{
Our algorithms are compatible with computing the factorized representation of the output, which has the potential of reducing the communication and storage cost to strictly less
than $OUT$ for many instances. There is a trade-off of course, as query computation over
factorized representations becomes more complicated. While exploring this trade-off is beyond the scope of this paper, we note this additional potential of the TAG encoding and the 
intriguing avenue it suggests for future work.}.

In summary, if we desire the collection of the join output in unfactorized representation,
we require the total communication $O(OUT + min(IN,OUT))$
and the total computation in $O(IN + OUT + min(IN,OUT))$ $\subseteq O(IN + OUT)$.
To leave the unfactorized join output distributed across vertices requires
communication cost in $O(min(IN,OUT)$ and computation cost in $O(IN + OUT + min(IN,OUT))
\subseteq O(IN + OUT)$.
To collect the factorized join output, the algorithm 
requires communication cost in $O(min(IN,OUT))$ and computation cost in 
$O(IN + min(IN,OUT)) \subseteq O(IN)$.
Finally, leaving the factorized join output distributed across vertices requires
communication cost in $O(min(IN,OUT))$ and computation cost in 
$O(IN + min(IN,OUT)) \subseteq O(IN)$.

A lax upper bound that covers all cases is therefore $O(IN + OUT)$ for both
computation and communication cost.

\paragraph{Dependence on the number of hardware processors.}
The vertex-centric model of computation makes the conceptual assumption
that each vertex is a processor. 
This is of course just an abstraction as the vertex processors are virtual, several of
them being simulated by the same hardware processor in practice.
Our complexity analysis is carried out on the abstract model, hence it overestimates
the communication cost by counting each message as inter-processor when a large number of messages are actually intra-processor
and do not tax the bandwidth of the interconnect. 

\subsubsection{Comparison to other algorithms}\label{sec:comparison-two-way}
Regarding both total communication and computation, our algorithm has the same lax upper bound $O(IN+OUT)$ as the computation upper bound of the classical sequential 
Yannakakis' algorithm \cite{yannakakis}, with the advantage of parallelism due to the vertex-centric nature.

State-of-the-art parallel join algorithms are mainly based on two techniques: hashing and sorting. These algorithms are usually based on the MPC \cite{koutris_1} and Map-Reduce \cite{afrati_1} computational models in a cluster setting. 

The parallel hash-join algorithm~\cite{koutris_1}
is the most common approach used in practice, where tuples are distributed by hashing on the join attribute value.
It achieves the same total communication complexity of $O(IN+OUT)$, assuming the 
unfactorized result from each processor is collected in a centralized location. 
In the scenario of a distributed, factorized representation of the join result, 
parallel hash-join requires communication $O(IN)$ while our vertex-centric join requires communication in $O(min(IN,OUT))$, which is better when the join is selective
(the computation cost is the same, $O(IN)$).


The parallel sort-join algorithm \cite{hu_1} is an MPC-based algorithm designed to handle arbitrarily skewed data, and measures the communication complexity per processor. We do not consider processor load balance in the scope of this paper, leaving it for future work. For apples-to-apples comparison, we derive from~\cite{hu_1} the total communication cost as 
$O(IN + \sqrt{p \cdot OUT} + OUT)$,  where $p$ is the number of processors and the last term describes the cost of streaming the final output to a centralized location. 
A skew resilient generalized version of a parallel hash join is also presented in \cite{koutris_2, koutris_book}, and achieves the same total communication cost as parallel sort-join. 
Depending on the size of the output and how skewed the input is, parallel sort-join will require more total communication. Otherwise (skew-free input), it achieves the same total communication cost as our vertex-centric algorithm.
The parallel sort-join requires the input to be sorted on the join attribute via
a reshuffling phase (which we do not require).
Such sorting incurs a communication cost of $O(IN)$, but requires additional supersteps, degrading the parallel sort-join's performance on ad-hoc queries that join on different attributes and in scenarios where the data is not read-only.

In summary, using the vertex-centric BSP model we can compute a single-attribute
two-way join over the TAG representation of the input relations matching the communication complexity of the 
best-known parallel algorithms (and even improving on it for the distributed factorized 
output scenario), while saving the query-dependent reshuffling they each require (for hashing or sorting).

\subsection{Join on Multiple Attributes}\label{sec:multi_attribute_join}
\begin{figure}[]
  \centering
  \includegraphics[width=0.7\linewidth]{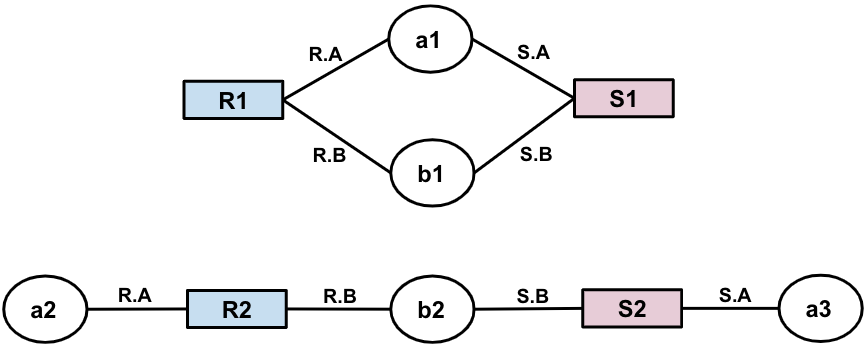}
  \caption{TAG instance for a two-way join on two attributes: R(A,B,C) $\Join$ S(A,B,D).}
  \label{two_join_more}
  \Description{}
\end{figure}
A necessary building block to generalize to arbitrary queries is to consider join conditions
on multiple attributes. We reuse the algorithm described in Section~\ref{sec:single_attrib_join} with a small adjustment in the reduction phase.
Specifically, the adjustment
concerns the part where a vertex corresponding to a join attribute checks whether it joins tuples from two relations or not. We illustrate the idea in the example below, and also explain the need for adjusting the algorithm.

\begin{newexample}
Consider the following query with natural join on two attributes: R(A,B,C) $\Join$ S(A,B,D).
The example input instance is shown in Figure \ref{two_join_more}.
If we apply the algorithm described in Section~\ref{sec:single_attrib_join} as is,
vertices corresponding to join attributes $A$ and $B$ check in parallel, independently of each other, whether they are join values (by iterating over their outgoing edges).
However, this results in incorrect output, e.g vertex $b2$ ends up joining two tuples $R2$ and $S2$ despite the fact that they disagree on their $A$ attribute values.
The problem stems from the fact that $A$-attribute and $B$-attribute vertices do not
communicate with each other.

Similar to the single-attribute join, we want to make only one join attribute resolve the join condition in the reduction phase.
Let's pick vertices of attribute $B$ for that role, and make tuple vertices send them their values of attribute $A$.
$B$-attribute vertices then perform an intersection of the received $A$ values from both sides, and computation proceeds further only for those values that survived the intersection.
Vertex $b1$ performs the following intersection $\{a1\} \cap \{a1\} = \{a1\}$, which succeeds for value $a1$ meaning that tuple vertices that sent this value join on $(a1,b1)$. 
$b1$ therefore notifies tuple vertices $R1$ and $S1$.
But vertex $b2$ is eliminated by the reduction phase, since the intersection 
$\{a2\} \cap \{a3\} = \oslash$, so tuple vertices $R2$ and $S2$ are no longer activated
despite agreeing on the $B$ attribute.
The ensuing collection phase runs unchanged according to Section~\ref{sec:single_attrib_join}.
\hfill$\Box$
\end{newexample}

We reduce a join on two attributes $X_1$ and $X_2$
to a join on a single attribute by having tuple attributes
send their value of the $X_2$ attribute to the $X_1$-attribute vertices.
Each $X_1$-attribute vertex intersects these values and messages back only to the
tuple vertices whose $X_2$ value is in the intersection.

The two-attribute join generalizes to a multi-attribute join 
on $X_1,X_2,\ldots ,X_n$, by sending a message with $X_2,\ldots ,X_n$ attribute values to the coordinating $X_1$-attribute vertex. 
An alternative is to compute the intersection in $n$ stages. First, send the $X_2$ values to the $X_1$ attribute vertices. 
These notify the tuple vertices for which the intersection on $X_2$ succeeds. In turn,
at the next superstep they send the values of their $X_3$ attributes, etc. 
This prunes messaging more aggressively, in that values of $X_i$ attributes are sent only by tuple vertices that are certain to have a join partner with respect to the first $i-1$ attributes.

\subsubsection{Cost Analysis}
For the two-way join on multiple attributes, we obtain the same
$O(IN+OUT)$ complexity for both communication and computation as for the
single-attribute join.
The difference is in the complexity of Superstep (1),
which is no longer just $O(min(IN,OUT))$. This is because the attribute vertices
no longer communicate with tuple vertices that are guaranteed to have a join partner,
but rather with tuple vertices that join on the first attribute, yet may not join
on the others. Still, since the set of input tuples is disjointly partitioned by 
the first join attribute value, the total number of tuples receiving messages
cannot exceed the total number of tuples in the input, hence the communication
is upper bounded by $IN$.
In a new Superstep, attribute vertices then obtain two sets of $B$ values, 
$B_R$ and $B_S$. The tuples messaged to next are those that have join partners
with respect to both $A$ and $B$, and the number of messages is again upper bounded by 
$min(IN, OUT)$.
The total communication complexity of the reduction phase is therefore 
$O(IN + min(IN,OUT)) \subseteq O(IN)$.
For centralized unfactorized output, reduction plus collection yield $O(IN + OUT)$.

As for the computation complexity, vertex attributes compute the intersection of 
$B_R$ and $B_S$, which can be done in quasi-linear time using hashing, or
in $nlog\ n$ worst-case time (it is customary for parallel join literature to hide
a polylogarithmic factor in their complexity analysis using the $\tilde{O}$ notation).
Thus the computation cost of the reduction phase is $\tilde{O}(IN)$. 
The collection phase proceeds the same way as in a two-way join on a single attribute algorithm, hence the same cost analysis applies.
The above analysis applies unchanged to the $n$-attribute two-way join, yielding the same
upper bound because the number of messages sent by tuple vertices to the coordinating
attribute vertices is the same (they are tuples, their width is admittedly larger but
bounded by the schema size and independent of the data size).

\section{Acyclic Multi-way Joins}\label{sec:acyclic_multiway}
We extend the two-way join algorithm to multi-way joins, as long as they are acyclic. 
For presentation simplicity, our treatment is confined to single-attribute
joins, with the understanding that these can be generalized to multi-attribute joins
as described in Section~\ref{sec:multi_attribute_join}.

\subsection{TAG Traversal Plan}\label{sec:tag_plan}
\begin{figure}[]
  \centering
  \includegraphics[width=0.7\linewidth]{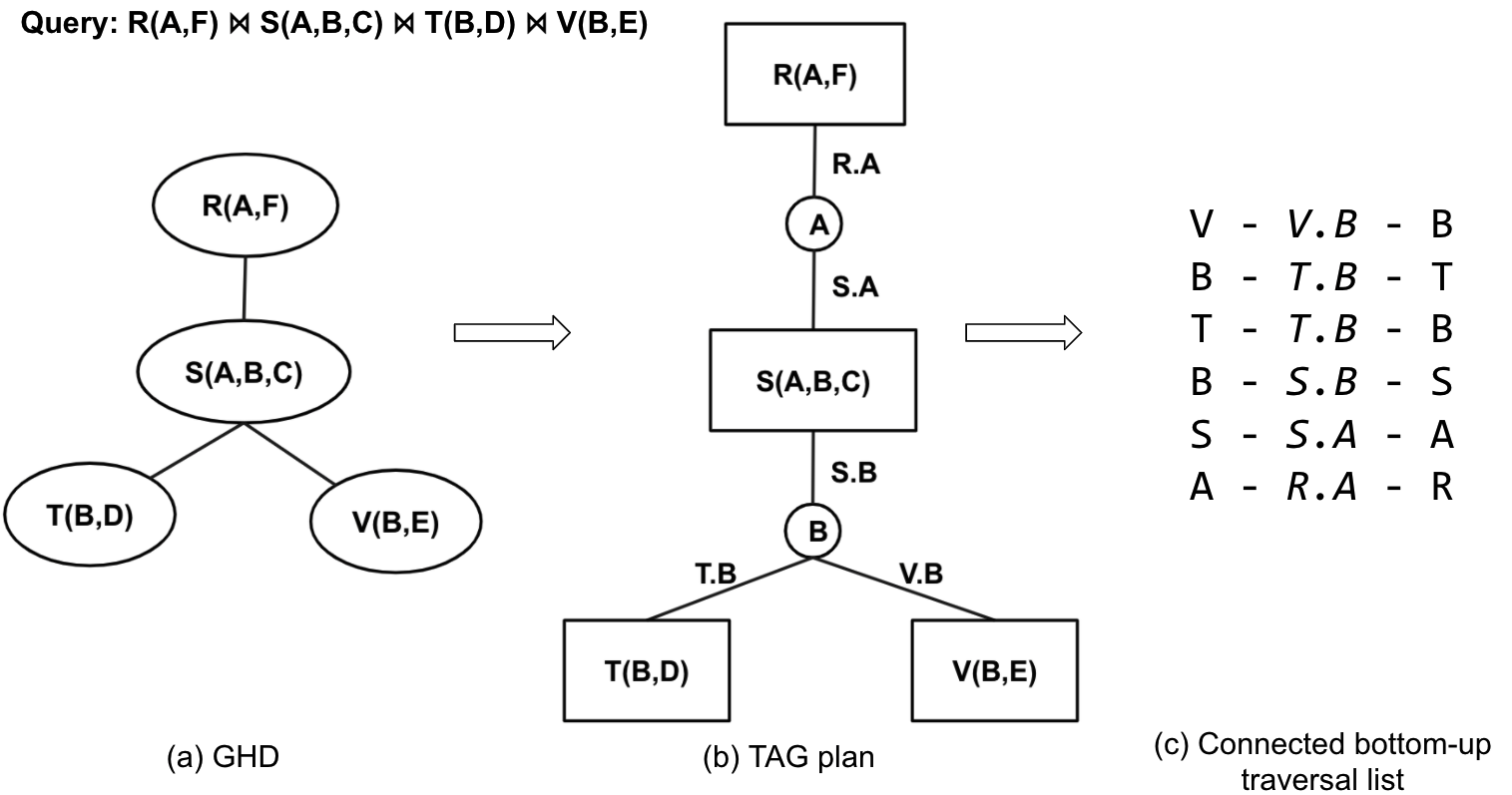}
  \caption{Example of a query plan translation}
  \label{query_plan}
  \Description{}
\end{figure}
We use a generalized hypertree decomposition (GHD) of the join query as a basis to obtain a TAG traversal plan (TAG plan), which in turn will help drive the vertex program.

We refresh the notion of GHD, referring to \cite{gottlob_1,gottlob_2}
for a more detailed treatment. 
A {\em GHD}
is a tree decomposition of a query, such that each node of the tree
(referred to as a 'bag') is assigned a set of attributes $\mathcal{A}$ 
and a set of relation names $\mathcal{R}$. For every bag, the schema of each
$R\in\mathcal{R}$ is included in $\mathcal{A}$. Moreover, every relation mentioned
in the query is assigned to some bag. Finally, for any attribute $A$,
all bags $A$ occurs in form a connected subtree.
A GHD where each bag is labeled by a single relation is called a {\em join tree}. 
Recall that a query is {\em acyclic} if and only if it has a join tree \cite{beeri}. 

\paragraph{TAG plan.} The TAG plan is itself a tree structure, constructed as follows from the join tree:

\begin{enumerate}
    \item For each bag, create a node labeled with the same relation.
    \item For each join attribute $A$ ($A$ occurs in at least two bags), create a node labeled $A$ if it does not exist already. 
    \item For each join attribute $A$, let $\mathcal{B}_A$ denote the set of all bags containing $A$. For each bag $b \in \mathcal{B}_A$, denote with $R_b$ the relation
    labeling $b$. Add an edge that connects the TAG plan node corresponding to $A$ with the TAG plan node corresponding to $b$. Label the edge with $R_b.A$.
\end{enumerate}
We call the elements of the TAG plan {\em nodes}, to avoid confusion with elements
of the TAG representation, which we call {\em vertices}.

\begin{newexample}
Figure \ref{query_plan}(b) shows the TAG plan constructed from the join tree in Figure~\ref{query_plan}(a).
\hfill$\Box$
\end{newexample}
Note that the example query only computes joins, thus it suffices to create plan nodes only for the join attributes. However, 
depending on the input query other attributes may be necessary for the computation (e.g. $GROUP$ $BY$ attributes and attributes
occurring in the $WHERE$ clause). In such cases we create plan nodes for these attributes as well.

\begin{algorithm}[h!]
\caption{GenSteps: Generate a list of traversal steps}
\label{alg:vc_program}
\KwIn{TAG plan $T = (V,E)$}
\KwOut{A list of TAG plan edge labels}
\KwInit

\tcp{a stack to store labels}
$labels \leftarrow$ empty stack\; 
\tcp{a dummy edge leading into root}
dummyEdge $\leftarrow (nil, nil, root(T))$\;
DFS(dummyEdge)\;

\algorithmicreturn{\ steps}\;
\
\SetKwFunction{FDFS}{DFS}
\SetKwProg{Fn}{Function}{:}{}
  \Fn{\FDFS{edge $inEdge$}}{
        \If{$t$ is not root of $T$}{
            $labels.push(inEdge.label)$;
        }
        
        \ForEach{$outEdge \in outEdges(v)$}{
            DFS($outEdge$);
        }
        
        \If{$v$ is not on rightmost root-leaf path in $T$} {
            $labels.push(inEdge.label)$;
        }
  }
\end{algorithm}

We next translate a TAG plan into a list $L$ of edge labels. 
Intuitively, $L$ is used to drive a vertex-centric program as follows: 
at superstep $i$, the active vertices send messages along their outgoing edges labeled $L(i)$.
We detail the list-driven vertex-centric program in Algorithm~\ref{alg:acyclic_join} below. First we explain how the list is 
generated by Algorithm \ref{alg:vc_program}.  

\paragraph{Connected Bottom-up Traversal} The list corresponds to what we call a 
{\em connected bottom-up traversal} of the TAG plan. The traversal is bottom-up in that it 
starts from the rightmost leaf and makes its way towards the root, eventually visiting
the entire plan, while first visiting all subtrees of a node $n$ before moving to $n$'s parent.
Moreover, the traversal is {\em connected} in the sense that each traversal step must
start from the node reached in the previous step. 
It should also be noted that reversing the list corresponds to a top-down (preorder) traversal of the TAG plan.

Algorithm~\ref{alg:vc_program} is implemented as a recursive DFS traversal of the input TAG plan tree. Each call of function DFS visits a TAG tree node $n$ by taking as input the incoming edge from $n$'s parent to $n$ and next it proceeds recursively to $n$'s children along $n$'s outgoing edges. 
Since the root of the TAG tree has no incoming edges, we start the traversal from a dummy edge leading into the root
and specifying no source or label (line 2).
Whenever we reach node $n$ from its parent, we record the incoming edge label (line 7). When returning to the parent,
we record this label again (line 11). The steps are stored in LIFO order using a stack, so that the first step in the sequence corresponds to the in-edge of the very last leaf node visited, and the successive steps correspond to moving up the TAG tree from there to the root, in connected bottom-up discipline.

\begin{newexample}
See Figure \ref{query_plan} (c) for the list of labels obtained from the TAG plan in Figure~\ref{query_plan} (b). Notice that it corresponds to 
the connected bottom-up traversal of the plan starting from the rightmost 
leaf (labeled $V$). 
For convenience, we
show to the left and right of each edge label the source, respectively destination of the traversal step.
\hfill$\Box$
\end{newexample}

\subsection{Vertex-Centric Algorithm}
\begin{algorithm}
\caption{Vertex Program}
\label{alg:acyclic_join}
\KwIn{$startLabel$: a vertex label}
\KwIn{$labels$: a stack of edge labels}
\SetKwFor{ForEach}{foreach}{do in parallel}

$ActiveVertexSet \leftarrow$ \{all vertices labeled $startLabel$\}\;
$direction \leftarrow$ UP \tcp*[l]{bottom-up traversal first} 
\textbf{Reduction Phase:}\\
\While{ $labels$ not empty }{
$currentLabel \leftarrow labels.pop()$\;
$reverseOrder.push(currentLabel)$\; 

\ForEach{$v\in ActiveVertexSet$}{
    \For{each $id \in v.incomingMsgQueue$}{
        insert $id$ into $v.markedEdges$\;
    }

 \If{$direction$ = UP}{
 \For{each $e \in outEdges(v)$}{
    \If{$e.label = currentLabel$}{
        send message $v.id$ to $e$'s target\; 
     }
 }
 }
 \If{$direction$ = DOWN}{
 \For{each $e \in outEdges(v)$}{
    $t \leftarrow$ target of $e$\;
     \If{$e.label = currentLabel$ AND $t \in v.markedEdges$}{
        send message $v.id$ to $t$\;
        update $v.markedEdges$\;
     }
 }
 }
}

\If{$labels$ is empty AND $direction$ = UP}{
  $labels \leftarrow reverseOrder$\; $reverseOrder.clear()$\;
  $direction \leftarrow$ DOWN\;
}

\tcp{synchronization barrier}
wait for all vertices to receive their messages\;
ActiveVertexSet $\leftarrow$ \{all message recipients\}\;
}

\textbf{Collection Phase:}\\
$labels \leftarrow reverseOrder$ \tcp{bottom-up order again}
\While{ $labels$ not empty }{
$currentLabel \leftarrow labels.pop()$\;
\ForEach{$v \in$ ActiveVertexSet}{
    $v.value \leftarrow$ join all tables in $v.incomingMsgQueue$\;
    \If{$v$ is a tuple vertex}{
        \eIf{this is first superstep in collection phase}{
            $v.value \leftarrow \{v.data\}$\;
        } {
            $v.value \leftarrow v.value \Join \{v.data\}$\;
        }
    }

    \For{each $e \in outEdges(v)$}{
        $t \leftarrow$ target of $e$\;
        \If{$e.label = currentLabel$ AND $t \in v.markedEdges$}{
            send message $v.value$ to $t$\;
        }
    }
    \If{$labels$  not empty}{
    \textbf{Output} $v.value$ \tcp*[l]{computation is done}
    }
}
\tcp{synchronization barrier}
wait for all vertices to receive their messages\;
ActiveVertexSet $\leftarrow$ \{all message recipients\}\;
}
\end{algorithm}

\paragraph{Vertex Program} 
The join algorithm performs a vertex-centric \revise{analogy} to Yannakakis' semijoin reduction technique in the following sense.
All vertices execute in parallel a vertex program comprising two phases: an initial reduction phase followed by a collection phase. 
The role of the reduction phase is to mark precisely the edges that connect 
the tuple and attribute vertices that contribute to the join.
The collection phase then traverses the marked subgraph to collect the actual join result.

Before detailing the logic implemented by the program, 
we note that its structure conforms to that of the classical vertex-centric BSP program.
Each iteration of the while loop starting at line 5 in Algorithm \ref{alg:acyclic_join}
implements a superstep. In each superstep, all active vertices compute in parallel, carrying out a computation and a communication stage. 
In the computation stage, each vertex processes its incoming messages (lines 8-9).
In the communication stage, each vertex sends messages to its neighbors via the edges whose label is dictated by variable $currentLabel$ (lines 11-13, 15-18). 
At the end of the superstep, vertices wait at a synchronization barrier for all messages to be received (line 24).
Only message recipients are activated for the next superstep (line 25).
Once all vertices finish processing the current traversal step, 
the edge label indicating the next traversal step is popped from the stack (line 5) and a new superstep is carried out. The reason why the input 
$labels$ must correspond to a connected traversal becomes apparent now:
the vertex-centric model requires each superstep to be carried out by the vertices activated by the previous superstep.

\paragraph{Reduction Phase}
The vertex program takes as input $startLabel$, the label of the rightmost leaf in the TAG plan $T$, and a list of edge labels (given in the stack $labels$), corresponding to the
connected bottom-up traversal of $T$.
The program starts by activating the tuple vertices labeled with $startLabel$ (line 1).
Next, it iteratively performs supersteps, one for every edge label in the stack. 
We say that the program is {\em driven by} $labels$.

The intuition behind the reduction phase is the following.
Recall that our TAG graph is bipartite: it consists of two kinds of vertices (attribute and tuple) and edges always go between the two vertex kinds. Therefore, the active vertex set alternates between containing exclusively
attribute vertices and exclusively tuple vertices. 
At every step, the active vertex set can be regarded
as a distributed relation (with the set of attribute vertices corresponding to a 
single-column table).
A superstep that starts from a set $S$ of active tuple vertices labeled $R$ and
activates next the attribute vertices reachable via edges labeled $R.A$ corresponds to taking the duplicate-eliminating projection on $A$ of
the $R$-tuples in $S$, $\pi_A(S)$, in the sense that the values in this column correspond precisely to the newly activated attribute vertices.
Conversely, a superstep that starts from set $S$ of active attribute vertices and next activates their neighbors reachable via edges 
labeled $R.A$ activates precisely the tuple vertices corresponding to the semijoin of table $R$ with an $A$ column of values from $S$, 
$R \lJoin \pi_A(S)$. The order in which the supersteps are performed by the reduction phase 
leads to a sequence of column projection and semijoins operations that corresponds to a Yannakakis-style reducer program.

The reduction phase starts with a connected bottom-up pass due to the $direction$ variable being initialized to $UP$ (line 2). 
At each superstep, vertices send their id to their neighbors along edges labeled by the 
current step (lines 12-13). Each vertex $v$ identifies its incoming join-relevant 
edges $e$ by their source id (received as the message payload). 
Vertex $v$ marks $e$ in its local state $v.markedEdges$ (line 9). 
Notice that if $v$ is a tuple vertex, it records which attribute vertex $a$ witnesses 
the fact that the tuple contributes to the join via $a$. 
If $v$ is an attribute vertex, it records which tuple vertex $t$ witnesses that $a$ contributes to the join.
The effect of the reduction pass is formalized by Lemma~\ref{l:reduction-up}. 

\begin{lemma}
\label{l:reduction-up}
Consider a list $L$ of edge labels, yielded by the connected bottom-up traversal of a TAG tree, and the execution of a vertex program driven by
$L$.
Let $T.A$ be the label at position $i$ in the concatenation of $L$ with its reverse. 
Denote with $R_i$ the distributed relation
corresponding to the vertex set activated by superstep $i$.
If $i$ is odd, then $R_i = \pi_A(R_{i-1})$.
If $i$ is even, then $R_i = T \lJoin R_{i-1}$. \hfill$\Box$
\end{lemma}

\begin{newexample}\label{ex:reduction-up}
Recall the edge label list shown in Figure~\ref{query_plan}(c). With the notation from Lemma~\ref{l:reduction-up},
the list induces the following sequence of operations: 
$R_0 := V, R_1 := \pi_B(R_0), R_2 := T \lJoin R1, R3 := \pi_B(R_2), R_4 := S \lJoin R_3, R_5 := \pi_A(R_4), R_6 := R \lJoin R_5$.
Notice that this sequence is a full reducer for table $R$, i.e. the last relation computed, $R_6$, 
contains precisely the $R$-tuples that participate in the multi-way join specified by the join tree in Figure~\ref{query_plan}(a).
\hfill$\Box$
\end{newexample}

As usual in full reducer programs, the bottom-up pass only reduces fully the root 
relation of the join tree (in Example~\ref{ex:reduction-up} that would be table $R$, whose full reduction is computed as $R_6$).
The other relations do not yet reflect the reduction of their ancestors in the join tree. To remedy this, one needs to apply further
semijoin reduction operations in an order given by a top-down traversal of the join tree.
In our vertex-centric adaptation, the vertex program switches to top-down mode  (lines 20-23).
It uses the $reverseOrder$ stack to reverse the order of input steps, thus obtaining a top-down order (line 6 and 21). 
The resulting top-down pass continues to apply reduction steps just like the bottom-up pass.

The only difference between the top-down and the bottom-up reduction
is that when sending messages, the former does not only check the edge label but it also makes sure to only signal via those edges that have been marked during the bottom-up pass (line 17). This ensures that the top-down pass never visits vertices that were already reduced away by the bottom-up pass.
The set of marked edges is updated further during the top-down pass to only include edges that are on the join result path (line 19).
Marked edges then guide the collection phase (\revise{line 39}), ensuring that it visits only the marked subgraph corresponding to the fully reduced relations.

\begin{newexample}\label{ex:reduction-down}
Continuing Example~\ref{ex:reduction-up}, the vertex program switches into the DOWN
pass, being driven by the reversed edge label list $R.A, S.A, S.B, T.B, T.B, V.B$, which
determines the sequence of operations
$R_7 := \pi_A(R_6), R_8 := S \lJoin R_7, R_9 := \pi_B(R_8), R_{10} := T \lJoin R_9,
R_{11} := \pi_B(R_{10}), R_{12} := V \lJoin R_{11}$ which fully reduces all tables. 
\hfill$\Box$
\end{newexample}

\paragraph{Collection Phase}
For the collection phase we reverse the order of steps once more, obtaining a bottom-up pass that starts from the vertices activated last 
by the reduction phase. During this phase, messages hold tables that correspond to intermediate results of the desired join.
In the computation stage, each vertex computes a value \revise{$v.value$} that corresponds to the join of the tables it receives via messages (line 31).
\revise{This value is a joined partial table and can be size-biased, its construction potentially creating load imbalance across vertices. This effect can be mitigated by keeping join results in factorized representation as long as possible (as discussed in our complexity analyses), 
but a full solution involves load balancing. Since we are targeting solutions on top of vertex-centric engines, we implicitly inherit their load balancing scheme and do not attempt to control it in this work. We believe that our encouraging experimental results render our approach interesting even before incorporating customized load balancing for vertex-centric SQL evaluation, which we leave for future work.
}
If the vertex is a tuple attribute, it also joins the computed value with the tuple it represents. This tuple is stored in the
$data$ attribute (lines 32-36).
In the communication stage, vertices propagate the computed value further up via marked edges (lines 37-40) 
\footnote{As shown, Algorithm~\ref{alg:acyclic_join} computes a full (projection-free) join.
The vertex program is compatible with pushing
projections early: if needed, appropriate projections are carried out by each vertex as it computes its value.}. 

When the root of a plan is reached, the computation completes and each active vertex 
outputs its computed value (line 42). The join result is the union of values output by vertices.

Observe that, compared to the classical Yannakakis reduction, 
our algorithm's reduction phase is more eager.

\begin{newexample}
The classical, centralized bottom-up pass of the Yannakakis reduction applied to the join tree from 
Figure~\ref{query_plan}(a) yields the three-step reduction sequence 
$Y_1 := S \lJoin V, Y_2 := Y_1 \lJoin T, Y_3 := R \lJoin Y_2$.
\end{newexample}

This is due to the necessity to contiguously navigate to children and backtrack to the parent. 

\subsubsection{Cost Analysis}
We are interested in the {\em data complexity} of our algorithm, which treats the query size as a constant.
The computation of a vertex in the reduction phase involves iterating over its incoming message queue and performing constant-time computation on each message. 
Each active vertex sends messages only via its outgoing edges, so at each superstep the total number of messages is bounded by the number of out-edges in the graph, while the size of each message is constant
(we treat the size of vertex ids as fixed by the architecture, therefore the message size is fixed).
Since the size of the TAG graph is linear in the size of the input database, 
the total computation and communication of each superstep of the reduction phase is $O(IN)$.
Note that the number of supersteps is independent of the data, being linear in the query size.
Therefore, the total computation and communication complexity of the reduction phase remain $O(IN)$.

The collection phase involves the traversal of the reduced subgraph, which has size linear in the size $OUT$  of the join result.
Each vertex constructs tuples by joining its own data with the intermediate results received.
These tuples are sent as messages along edges.
There is no redundant tuple construction within and across vertices, 
so the overall computation and communication is in $O(OUT)$.

Combining the complexity of the reduction and collection stages,
we obtain $O(IN+OUT)$ for the overall communication and computation costs of our vertex-centric acyclic join.

\subsubsection{Comparison to other algorithms}
Reference \cite{koutris_book} describes a hash-based parallel version of 
Yannakakis' algorithm in a distributed setting (the MPC model) with the same communication complexity of $O(IN+OUT)$.
The main difference from our algorithm is that data needs to be reshuffled (re-distributed
among processors) in a query-dependent way
for each join operation.
In our vertex-centric join,
the graph representation of the input database is never reshuffled, regardless of the query. 

The generalization of parallel sort-join algorithm to any acyclic join is presented in \cite{hu_2}, and has a total communication cost of $O(IN + \sqrt{IN \cdot OUT})$.
This outperforms our algorithm and parallel hash-join when 
the join output blows up to be lager than the input,
but it is worse for selective joins and it is equivalent when
the query involves only PK-FK joins.

Both parallel sort-join and parallel hash-join rely on query-dependent reshuffling
(re-sorting and re-hashing) of the input, which again impairs the applicability to ad-hoc queries that join on different attributes and to scenarios where the data is not read-only.


\section{Arbitrary Equi-Join Queries}\label{sec:arbitrary}
In this section, we extend our algorithm to arbitrary equi-join queries.
We begin with the famous triangle query in Section~\ref{sec:triangle}
and we extend the algorithm to $n$-way cycle queries in Section~\ref{sec:cyclic}. Section~\ref{sec:cartesian_product} presents an algorithm to compute a Cartesian product. Finally, we describe the complete TAG-join algorithm in Section~\ref{sec:complete_alg}.

\subsection{Triangle Query}\label{sec:triangle}
\begin{figure}[]
  \centering
  \includegraphics[width=0.4\linewidth]{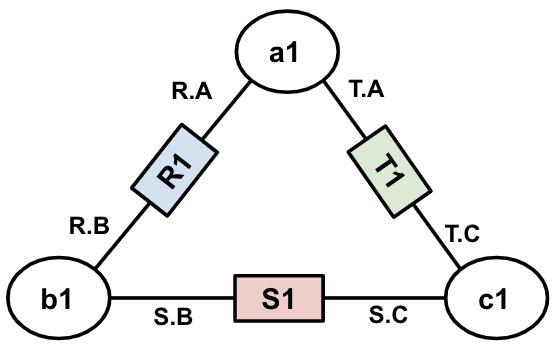}
  \caption{TAG instance for a triangle query in Example~\ref{ex:triangle}}
  \label{triangle}
  \Description{}
\end{figure}

We begin with the triangle query
\begin{displaymath}
R(A,B) \Join S(B,C) \Join T(C,A).
\end{displaymath}
First, we describe a first-cut vertex-centric triangle algorithm and show that it is optimal for a class of queries with primary-foreign key (PK-FK) join conditions.
We then improve upon it to achieve the complexity proportional to the worst-case output size defined by the renowned AGM bound \cite{agm}.
This is a tight bound that estimates query output size based on input relations cardinalities and structural property of a query called fractional edge cover ($\rho^*$). For a full review of AGM and fractional edge cover we refer to \cite{agm,grohe}.
The AGM bound for triangle query is $O(IN^{\frac{3}{2}})$, while the traditional RDBMS
plans with binary joins can run in time $O(IN^2)$ on some instances.
The AGM bound has led to the development of the class of worst-case optimal 
(centralized, sequential) join algorithms such as NPRR \cite{nprr}, Leapfrog Triejoin \cite{leapfrog} and Generic-Join \cite{skew_strikes}.
Our parallel vertex-centric algorithm's communication and computation complexities 
match the worst-case computational upper bound of these algorithms for triangle queries.

\subsubsection{Triangle Query for PK-FK Joins}\label{sec:trianglePKFK}
The main idea of the vertex-centric triangle algorithm is to start the computation from $A$-attribute vertices, and send their id values in both directions via paths that lead to $C$-attribute vertices such that if a cycle exists the values travelling from both sides meet at the final destination. Note that values propagated through the left side have a 
longer path to cross and pass through $B$-attribute vertices.
Any order of the traversal can be chosen, e.g. start from $B$-attribute vertices and propagate $B$ values in both directions to $A$-attribute vertices.    

\begin{newexample}\label{ex:triangle}
Let's illustrate the algorithm using the TAG instance in Figure \ref{triangle}. Attribute vertex $a1$ sends its id via edges $R.A$ (left side traversal) and $T.A$ (right side traversal).
Note that if vertex $a1$ \revise{has no incident $R.A$- or $T.A$-edge}, it deactivates itself. On the right side the value $a1$ travels along the edge $T.C$ to arrive at attribute $c1$. On the left side the value $a1$ arrives at the $c1$ vertex via the edge list $R.B, S.B, S.C$. The $c1$ vertex intersects messages received from the left side with the messages received from the right side, as described in Section~\ref{sec:multi_attribute_join}. Empty intersection, indicates that a vertex is not part of any triangle. Since the result of the intersection is $a1$, vertex $c1$ continues into the collection phase. To construct the output, $c1$ sends messages back following the trail of the $a1$ value to activate tuple vertices $S1$ and $T1$. Those tuples then send their values to $c1$, which then combines them and outputs a triangle $\{(a1,b1,c1)\}$.
\hfill$\Box$
\end{newexample}

\paragraph{PK-FK Optimality}
In the triangle query when a value reaches an attribute vertex it needs to be propagated further via its outgoing edges, i.e. every message gets replicated a number of times.
The replication rate is defined by the number of outgoing edges which is linear in the size of the input $IN$.
The number of messages that an attribute vertex can receive is also upper bounded by $IN$.
In the worst case the number of messages that need to be sent can blow up to $O(IN^2)$, which exceeds the worst-case AGM bound. However, the replication rate is not an issue with PK-FK joins, since the size of the PK-FK join result cannot exceed the size of the foreign key relation, which is at most $IN$. Assume the primary key of each input relation is its first attribute, e.g attribute $A$ is the primary key of relation $R$. The number of $A$ values that $b1$ can receive is at most $|R|$, but since $b1$ itself is a primary key value of $S$ tuple it can have at most one outgoing edge with label $S.B$. Thus, the number of messages sent by $b1$ is at most $|R|$. The same analysis applies to the $c1$ vertex.  

If all joins in the query are PK-FK joins, the  triangle query can be evaluated 
in the vertex-centric model with optimal communication and computation cost $O(IN + OUT)$, just like acyclic joins.

\subsubsection{Worst-case Optimal Triangle Query Algorithm}\label{alg:wcoj-triangle}
In order to match the worst-case optimal guarantees by keeping the complexity within the AGM bound we improve upon the algorithm above. We employ the strategy of the NPRR algorithm. NPRR splits the values of attribute $A$ in relation $R$ into heavy and light. 
A value $a$ is heavy if it occurs more times than a defined threshold value $\theta$ in relation $R$. Specifically, if $|\sigma_{A=a}R|$ > $\theta$ then (a,b) $\in R_{heavy}$, otherwise (a,b) $\in R_{light}$. 
As a result, the original triangle query can be decomposed as follows \cite{ngo2018}:
\begin{equation}
  (R_{heavy} \Join S) \lJoin T))  \cup ( (R_{light} \Join T) \lJoin S)
\end{equation}
We solve the triangle query separately for heavy and light cases and then union the results. We apply the vertex-centric triangle algorithm as described above, for simplicity we refer to it as vanilla triangle. The triangle algorithm proceeds as follows:
\begin{enumerate}
    \item \textbf{Initialization:} Activate $R$-tuple vertices and navigate to $A$-attribute vertices  via edge $R.A$. Each $A$-attribute vertex checks whether it's heavy or light. 
    \item \textbf{Heavy} $A$-attribute vertices execute vanilla triangle algorithm (as described in Example~\ref{ex:triangle}). 
    \item \textbf{Light} $A$-attribute vertices send "wake-up" messages to $B$-attribute vertices via their $R$-tuple vertex.  
    Activated $B$-attribute vertices then execute vanilla triangle algorithm to propagate their id values to $C$-attribute vertices.    
\end{enumerate}
Note that the number of outgoing edges with label $R.A$ indicate how many tuples in $R$ contain the current attribute vertex value, hence it becomes trivial for each $A$-attribute vertex to check whether it's heavy or light. The threshold value $\theta$ helps to bound the number of messages and not to exceed the AGM bound. In the heavy case, the replication of messages happens when $B$-attribute vertices send received heavy $a$ values via $S.B$-edges. This corresponds to the term $R_{heavy} \Join S$ in equation (1). The number of messages that can be received by $B$-attribute vertices is upper bounded by the total number of heavy $a$ values in the $R$ relation, which is at most $\frac{|R|}{\theta}$. Each message is sent via outgoing edge $S.B$. Since the total number of $S.B$ edges is $|S|$, 
the communication cost is $\frac{|R|}{\theta} \cdot |S|$ messages.
In the light case, the worst replication happens when $A$-attribute vertices send $b$ values via edge $T.A$, i.e. term $R_{light} \Join T$ in equation (1). The number of $b$ values that $A$-attribute vertices can receive is at most $\theta$, since all of these $b$ vertex values are connected to light $a$ vertices. Each $b$ value is sent via at most $|T|$ edges. This results in $\theta \cdot |T|$ total messages for the light case computation. 
Each vertex' computation is linear to the received message count.
Set $\theta = \sqrt{IN}$, where $IN$ defines the sizes of input relations, so that the complexity of the reduction phase is proportional to the AGM bound, i.e. $O(IN^\frac{3}{2})$. Taking into account that the size of the actual output is upper bounded by the AGM, the overall communication and computation cost of our triangle algorithm in a vertex-centric BSP model is $O(IN^\frac{3}{2})$.

Besides being sequential algorithms, both NPRR and Leapfrog Triejoin require expensive precomputation to build index structures for each input relation based on some attribute order. For example, the cost of indexing the input relations used in a given query is $O(n^2 \cdot IN)$, while indexing all the relations in the input database in advance to compute any query in the future is $O(n \cdot n! \cdot IN)$, where $n$ is the number of attributes \cite{nprr}. Our parallel triangle algorithm does not rely on any additional index structures to achieve the worst-case optimal guarantees.


\subsection{Cycle Queries}\label{sec:cyclic}
We obtain the evaluation of an $n$-way cycle as a straightforward generalization of the triangle join from Section~\ref{alg:wcoj-triangle}. Our algorithm goes through the same reduction and collection phases as defined in acyclic join algorithm.
It is known that we cannot obtain a full reducer for a cyclic query \cite{semijoin,semijoin2}, i.e. a sequence of semijoins to remove all dangling tuples. However, reduction still helps to eliminate some vertices that are not part of the output and marks the edges with additional information regarding cycles to guide the collection phase such that the number of constructed output tuples does not explode beyond worst-case output bound (AGM).
Given a cyclic query
\begin{displaymath}
R_1(X_1,X_2) \Join R_2(X_2,X_3) \Join \ldots \Join R_n(X_n,X_1)
\end{displaymath}
, the computation starts by activating $R_1$-tuple vertices and navigating to $X_1$-attribute vertices. 
Each $X_1$ attribute checks whether its heavy or light based on the predefined threshold value $\theta$.

The heavy $X_1$ vertices propagate their values via the edges on the left and right sides to be intersected at $X_{\ceil*{\frac{n}{2}+1}}$-attribute vertices. As messages get propagated, similarly to the reduction phase (bottom-up) in the acyclic join algorithm, each visited vertex marks the edges via which the messages are received. 
Note that the messages that are sent during the reduction phase contain two values: the $X_1$-attribute value and ID of the vertex that is sending this value. So, the marked edges info is extended to keep track of what $X_1$-attribute value is sent along each edge, i.e. each vertex can group marked edges info by the received $X_1$-attribute value.
Once the $X_1$-attribute values are intersected, $X_{\ceil*{\frac{n}{2}+1}}$-attribute vertices need to signal back which values are part of the cycle by sending those values via the marked edges again. This is similar to the top-down reduction phase in the acyclic join algorithm.
Recall, that each vertex maintains the marked edges info with respect to $X_1$ values, thus for $X_1$ values that form a cycle the edge markings get updated accordingly, and the rest are discarded (i.e. edge markings corresponding to $X_1$ values that did not survive the intersection).
Then collection phase proceeds to construct tuples and propagating the intermediate results further following the marked join edges with respect to $X_1$-attribute value that represents the start of the cycle. The computation for the heavy case completes at $X_{\ceil*{\frac{n}{2}+1}}$-attribute vertices. 
The light $X_1$ vertices send "wake-up" messages to $X_2$-attributes via $R_1$ tuple vertices. The activated $X_2$-attributes propagate their values now to be intersected at $X_{\ceil*{\frac{n}{2}+1}}$-attribute vertices. The rest is the same as for the heavy case, with the exception that marked edges info is organized with respect to $X_2$ values.

During reduction phase the messages are propagated via the paths on both sides until meeting in the middle. While for even cycles, the length of the paths on both sides is equal, in the case of odd cycles the path one side is longer than the other one. Thus, the number of messages sent along the longer path will dominate in the overall complexity. The number of heavy $X_1$ values is at most $\frac{|R_1|}{\theta}$, and the maximum replication rate is determined by the sizes of relations on the longer path, $\prod_{i=2}^{\ceil*{\frac{n}{2}}}|R_i|$.
The communication cost of reduction phase for heavy $X_1$ values is at most:
\begin{equation}\label{eq:heavy}
\frac{|R_1|}{\theta} \cdot \prod_{i=2}^{\ceil*{\frac{n}{2}}}|R_i|   
\end{equation}
For the light case, where the number of $X_2$ values that are propagated for the intersection is at most $\theta$, since each light $X_1$ is connected to at most $\theta$ $X_2$-attribute vertices via $R_1$ tuple, the maximum replication rate is defined as $\prod_{i=\ceil*{\frac{n}{2}+1}}^{n}|R_i|$.
Then communication cost of reduction phase for light $X_1$ values is at most:
\begin{equation}\label{eq:light}
\theta \cdot \prod_{i=\ceil*{\frac{n}{2}+1}}^{n}|R_i|   
\end{equation}
Note each vertex does computation that is linear in the size of the received communication, hence the total cost of reduction phase, including heavy and light stages, is the sum of the equations (\ref{eq:heavy}) and (\ref{eq:light}). 
It is important to set the threshold value $\theta$ in order to keep the cost of reduction phase within the AGM bound, the largest possible output size. The collection phase only sends messages via the edges marked during reduction phase, thus its communication cost is not going to exceed cost of reduction phase, i.e. also upper bounded by the AGM estimate.

Let the size of input relations be $IN$, then the total communication and computation cost of evaluating an $n$-way cycle query is
\begin{equation}\label{eq:total}
\frac{IN}{\theta} \cdot IN^{\ceil*{\frac{n}{2}-1}} + \theta \cdot IN^{\ceil*{\frac{n}{2}-1}}
\end{equation}
For example, recall that for triangle query, we set the $\theta = \sqrt{IN}$ that makes the total cost $IN^{\frac{3}{2}}$, which matches the established AGM bound for the triangle query.

\begin{newexample}\label{ex:cycle_query}
Consider a 5-way cycle. We begin by activating $R_1$-tuple vertices, and navigate to $X_1$-attribute vertices. Each active $X_1$-attribute vertex checks whether its heavy or light. The heavy $X_1$ vertices propagate their values in both directions via paths that lead to $X_4$-attribute vertices, marking the edges along the way. $X_4$-attributes intersect received values to figure whether there exists cycle or not, and for values that succeeded they send signals along the marked edges and confirm marking for join edges. Then collection phase proceeds to construct output tuples and send them to $X_4$-attribute vertices. The light $X_1$ vertices send "wake-up" messages to $X_2$ attribute vertices via $R_1$-tuple vertices. Activated $X_2$ vertices now proceed by propagating their id values via paths leading to $X_4$
-attribute vertices. The rest is the same as in the case of heavy values.

In the heavy case, the biggest replication of $X_1$ values happens on the path that goes through relations $R_2$ and $R_3$ before reaching $X_4$-attributes. Using equation (\ref{eq:heavy}) we know that the cost of heavy stage is $\frac{|R_1|}{\theta} \cdot |R_2| \cdot |R_3|$, where the first term denotes the number of heavy $X_1$ values.
In the light case, using equation (\ref{eq:light}) we estimate the cost as $\theta \cdot |R_4| \cdot |R_5|$. The maximum replication of $X_2$ values happens on the path going through $R_5$ and $R_4$. The total number of $X_2$ values that are connected to the light $X_1$ values in relation $R_1$ is at most $\theta$.
Setting the threshold value $\theta = \sqrt{IN}$, results in $O(IN^{\frac{5}{2}})$ complexity for both heavy and light $X_1$ values, i.e. proportional to the AGM bound of a 5-way cycle.
Collection phase cost is not going to exceed the worst-case output estimate, and so the total communication and computation cost of a vertex-centric 5-way cycle query is $O(IN^{\frac{5}{2}})$.  
\end{newexample}

Note that for an $n$-way cycle query with PK-FK join conditions only, we can use the same strategy as described in Section ~\ref{sec:trianglePKFK}, since the cost of the reduction phase is not going to exceed the size of the biggest relation in the cycle.

\subsection{Cartesian Product}\label{sec:cartesian_product}
In the vertex-centric BSP model a Cartesian product of relations can be computed via the communication with global aggregator vertex, whose ID is known to all vertices. Recall that the computational model allows vertices to send messages directly to any other vertex using the ID of that vertex.
Consider the Cartesian product between relations $R(A,B) \Join S(C,D)$. Lets define a global aggregator vertex as $GA$.

\textbf{Algorithm A.} A naive algorithm sends the data of tuple vertices $R$ and $S$ to the aggregator $GA$.
Then $GA$ combines the received tuples to compute the Cartesian product. The communication cost is linear in the sizes of the input relations, O($|R|$+$|S|$), and the computation cost to construct the output tuples is exactly the size of the output, $O(|R| \cdot |S|)$.  
So, the total cost of computing the Cartesian product is $O(|R| \cdot |S|)$.
However, the computation is mostly sequential and does not take advantage of the parallelism of the vertex-centric model, i.e. multiple vertices computing in parallel. 

\textbf{Algorithm B.} A better algorithm exploits the parallelism of the underlying computational model and computes the Cartesian product in a distributed way, but requires extra rounds of computation and communication.
The idea is to forward the tuples of $S$ to $R$-tuple vertices. In order for $S$-tuple vertices to communicate directly with $R$-tuple vertices, they need to know the IDs of $R$-tuple vertices.
The computation starts by activating $R$ and $S$ vertices. The active tuple vertices send their IDs to the global aggregator $GA$, which then transmits the IDs of all $R$ vertices to each $S$-tuple vertex.
Each $S$-tuple vertex sends its tuple data as a message to all $R$-tuple vertices. Each $R$-tuple vertex receives $|S|$ messages, and combines each message with its own tuple data to construct the output tuples. The result of the Cartesian product is now distributed among $R$-tuple vertices.

The aggregator vertex $GA$ receives at most $|R| + |S|$ messages, and sends each received ID of $R$-tuple vertex $|S|$ times, which incurs at most $O(|R| \cdot |S|)$ communication and computation cost.
Note that $GA$ is working with messages containing $ID$ values only, the size of ID is smaller compared to the size of the entire tuple.
Each $S$-tuple vertex is going to send $|R|$ messages with its tuple data, resulting in the communication and computation cost of $O(|R| \cdot |S|)$ over all $S$-tuple vertices.
Each $R$-tuple vertex receives $|S|
$ messages and computes its part of the final result. Thus, the total computation and communication cost of the algorithm is $O(|R| \cdot |S|)$, i.e. does not exceed the actual size of the Cartesian product result.

Either of the algorithms described in the above can be extended to the Cartesian product of $n$ relations. Although, we focus on \textbf{Algorithm B} in the remainder of the section.
The idea is to forward the tuples of $n-1$ relations to the tuple vertices corresponding to the $n^{th}$ relation, given that the IDs of vertices corresponding to the $n^{th}$ relation are communicated to the other tuple vertices using global aggregator $GA$. Observe, that we can either send all the data to the tuple vertices of the $n^{th}$ relation at once, and let those tuple vertices compute the final result. Or, instead do a cascade of binary Cartesian products applying \textbf{Algorithm B} on each. The underlying vertex-centric BSP model, allows us to do multiple binary Cartesian products in parallel.  \begin{newexample}
Lets compute the Cartesian product of 4 relations $R_1$, $R_2$, $R_3$ and $R_4$. First, by applying Algorithm B on $R_{1,2} = R_1 \times R_2$ and $R_{3,4} = R_3 \times R_4$, we compute two binary Cartesian products in parallel. We store $R_{1,2}$ tuples at $R_1$-tuple vertices, and $R_{3,4}$ tuples at $R_3$-tuple vertices, where $|R_{1,2}| = |R_1| \cdot |R_2|$ and $|R_{3,4}| = |R_3| \cdot |R_4|$. The complexity of this stage is determined by $O(max(|R_1| \cdot |R_2|, |R_3| \cdot |R_4|))$.

Then, we apply Algorithm B on two intermediate results to compute the final output. We forward $R_{3,4}$ tuples stored at $R_3$-tuple vertices to $R_1$-tuple vertices. Each $R_1$-tuple vertex combines the received $R_{3,4}$ tuples with $R_{1,2}$ tuples that are stored locally. The total communication and computation cost is therefore $O(|R_{1,2}| \cdot |R_{3,4}|)$, which is equal to $O(|R_1| \cdot |R_2| \cdot |R_3| \cdot |R_4|)$.
\end{newexample}

\subsection{TAG-join Algorithm}\label{sec:complete_alg}
By combining acyclic and cyclic join strategies (as well as the Cartesian product) we obtain a complete TAG-join algorithm to evaluate an arbitrary equi-join query.
Given a generalized hypertree decomposition of a query we construct a corresponding TAG plan, and evaluate a query in two steps:
\begin{enumerate}
    \item \textbf{Compute intermediate results.} 
    Evaluate each subquery corresponding to a bag of a tree decomposition~\footnote{Bags that contain more than one relation.}.
    \item \textbf{Compute acyclic join.} Run multi-way acyclic join algorithm to get the final result.
\end{enumerate}
We next present the algorithm steps in detail and analyze it formally. The main result of TAG-join is stated by the theorem below. 
\begin{theorem}
\label{th:main}
Given any equi-join query and its GHD with a factional hypertree width $w$, a vertex-centric TAG-join algorithm can compute the query with $O(IN^w+OUT)$ communication and computation cost.  
\end{theorem}

\subsubsection{Step (1): Compute Intermediate Results}\label{sec:computeIR}
We now show that acyclic, cyclic and Cartesian product algorithms are sufficient to evaluate any subquery in a tree decomposition bag with complexity proportional to the AGM bound of a subquery (bag).
We begin with the special case of queries when each relation has at most 2 attributes, and then extend the result to a general case of multi-attribute relations.

Atserias, Grohe and Marx (AGM \cite{agm}) established a tight bound on the maximum possible query result size using optimal fractional edge cover of a query. Given a join query (i.e. hypergraph of the underlying query) an edge cover is a minimum number of relations (hyperedges) such that each attribute (vertex) is contained in at least one relation. The minimum edge cover problem can be formulated as a linear program, where a feasible solution is a set of a non-negative weights assigned to each relation, such that each attribute is covered by the total weight of at least 1. 
If a relation is included in a minimum edge cover then its weight is 1, otherwise it is assigned 0.
A fractional edge cover is a relaxation of the integer linear program of the edge cover, where solution is a set of rational non-negative weights.
A fractional edge cover number (the sum of all weights) is the minimum among all possible fractional edge cover solutions of the query.

Balinski \cite{balinski} showed on a graph structure that a feasible solution to the fractional edge cover linear problem has half-integral values $0, \frac{1}{2}$ or $1$. The proof of this half-integrality property from \cite{schrijver} is then adapted to a relational join setting in \cite{nprr}.
\begin{lemma}[\textbf{Half-integrality lemma from~\cite{nprr}}]
\label{lem:half-integral}
Given a join query, where all input relations are binary, let \textbf{$e$} be the fractional edge cover solution, where each weight $e_i \in \{0, \frac{1}{2}, 1\}$. Then, the set of relations with $e_i = 1$ form a union of stars. And the set of relations with $e_i = \frac{1}{2}$ form a collection of odd-length vertex-disjoint cycles. The collection of cycles are also vertex-disjoint from the union of stars.  
\end{lemma}

\begin{figure}[]
  \centering
  \includegraphics[width=0.6\linewidth]{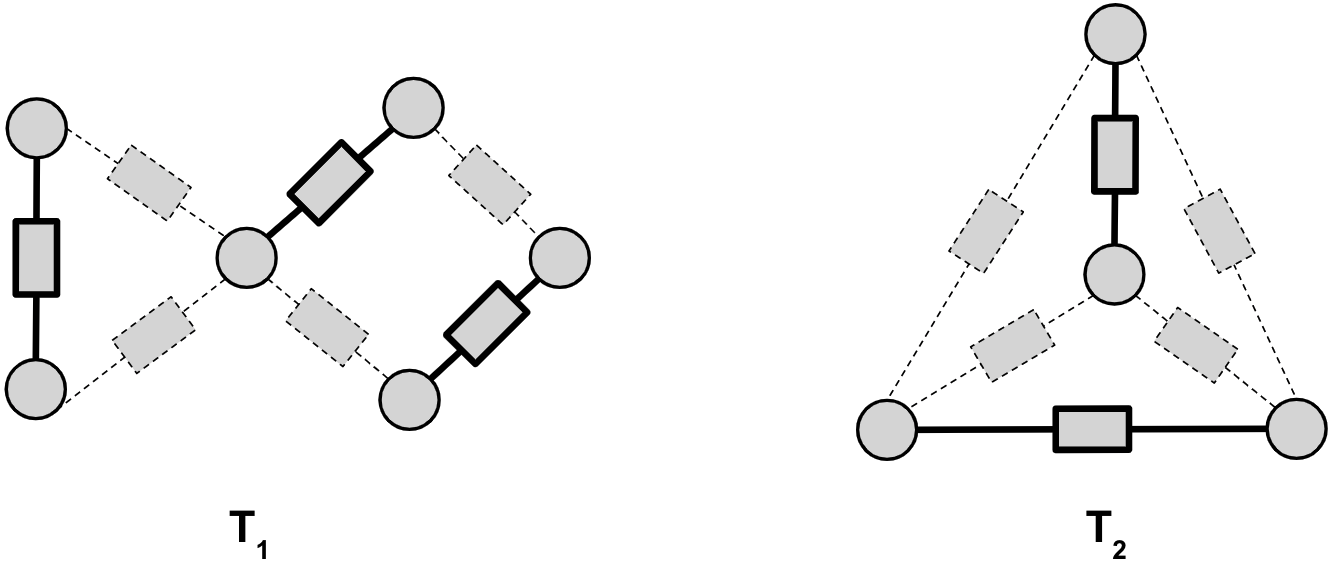}
  \caption{TAG plans of queries that form a union of stars. Each star is shown in bold solid lines. }
  \label{decomposition_1}
\end{figure}

\begin{figure}[h!]
  \centering
  \includegraphics[width=0.6\linewidth]{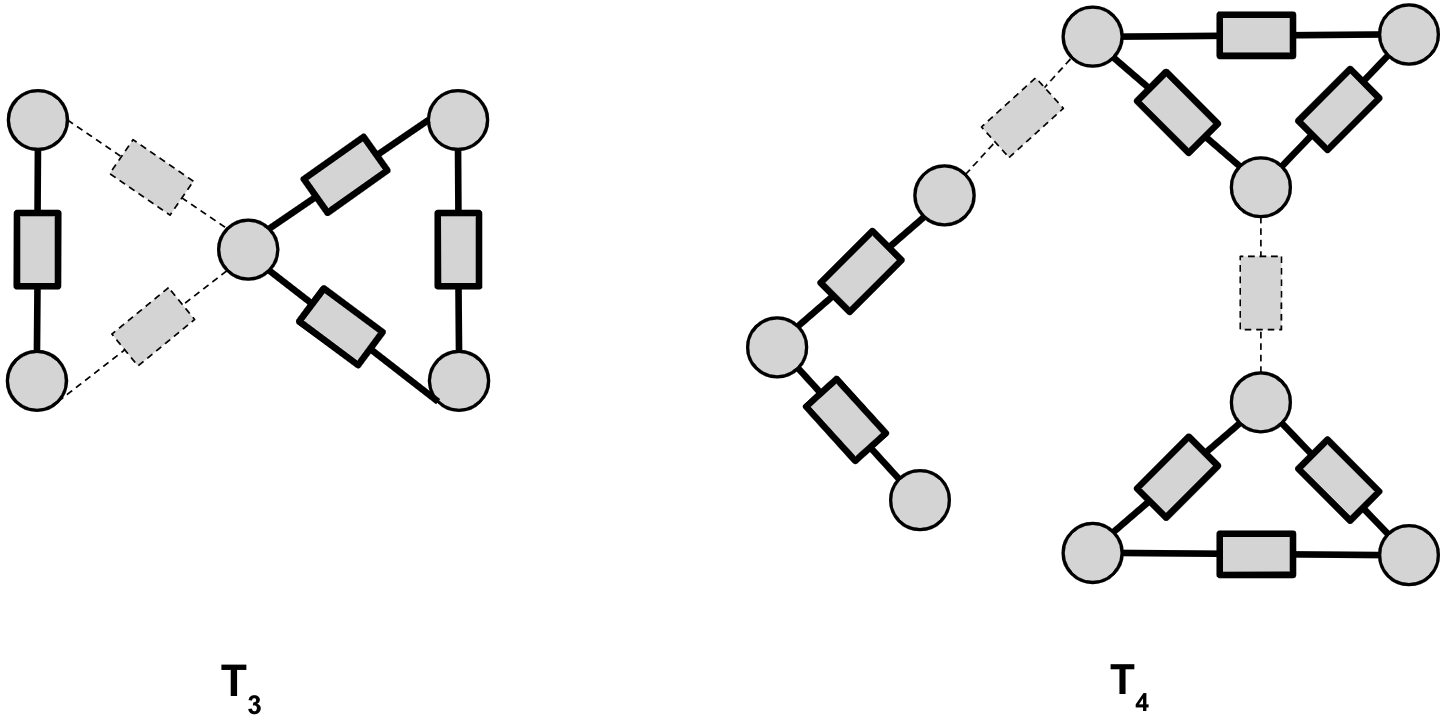}
  \caption{TAG plans of queries that form stars and cycles. }
  \label{decomposition_2}
\end{figure}

Refer to \cite{nprr} for the detailed proof of this lemma.
Figure~\ref{decomposition_1} shows examples of queries that each form a union of stars based on their fractional edge cover solutions.
Query with the corresponding TAG plan $T_1$ has a feasible edge cover solution $(1,0,0,1,0,1,0)$. Each relation that is assigned weight $1$ is shown in bold solid lines, and each represent a star. TAG plan $T_2$ corresponds to a 4-clique query, which can also be defined as a union of two stars.
Figure~\ref{decomposition_2} depicts TAG plan $T_3$, which is decomposed into triangle and star components after the application of the half-integrality lemma. And TAG plan $T_4$ is split into a collection of two cycles and a star. 

Following Lemma~\ref{lem:half-integral}, let $C$ be the collection of odd-length cycles and $S$ be the union of stars,
then \cite{nprr} showed that the worst-case output size (AGM bound) is
\begin{equation}
\label{eq:half_integral}
\prod_{i=1}^{n}|R_i|^{e_i} = (\prod_{R_i \in S}|R_i|) \cdot \prod_{c \in C} \sqrt{\prod_{R_i \in c} |R_i|},
\end{equation}
where $n$ is the number of relations in a query.
Consequently, any join query on binary relations
can be computed with cost proportional to the AGM bound of the query.

We apply Lemma~\ref{lem:half-integral} to split each subquery associated with a bag of a GHD into a collection of disjoint cycles and/or a union of stars. We proceed as follows:
\begin{itemize}
    \item Each odd-length cycle in the collection is evaluated using the algorithm described in Section~\ref{sec:cyclic}.
    \item A star is a special case of an acyclic query, therefore we can apply the acyclic join algorithm from Section~\ref{sec:acyclic_multiway}. And the union of stars $S$ is computed using the Cartesian product algorithm presented in Section~\ref{sec:cartesian_product}.
    \item Cycles and/or the union of stars are then joined by applying the Cartesian product algorithm.
    \item The Cartesian product results also need to be (semi-)joined with the remaining relations (i.e where $e_i = 0$) using an acyclic join algorithm \footnote{Performed after any Cartesian product to eliminate "dangling tuples" as early as possible.}.
\end{itemize}

\begin{lemma}
\label{lem:ir_cost}
Let a query be defined as a collection of disjoint cycles and/or a union of stars given its fractional edge cover solution $e = (e_1, \ldots ,e_n)$. Then it can be computed in the vertex-centric BSP model with communication and computation cost proportional to the worst-case output size of the query, $O(\prod_{i=1}^{n}|R_i|^{e_i})$. 
\end{lemma}
\noindent
\emph{Proof.} 
It follows from Section~\ref{sec:cyclic} that the cost of computing an odd-length cycle is $O(\prod_{R_i \in c} |R_i|^{\frac{1}{2}}) = O( \sqrt{\prod_{R_i \in c} |R_i|})$.
Applying the result of Section~\ref{sec:cartesian_product} on the Cartesian product, the cost of computing the union of stars is $O(\prod_{R_i \in S}|R_i|)$.
The results of cycles and stars are combined using the Cartesian product, thus we again take the product of the terms corresponding to the union of stars ($S$) and the cycles ($C$): $(\prod_{R_i \in S}|R_i|) \cdot \prod_{c \in C} \sqrt{\prod_{R_i \in c} |R_i|} $.
This cost dominates the cost of (semi-)joining with the remaining relations where weight is equal to 0.
As a result we obtain the total communication and computation cost proportional to the worst-case output estimate of the query as shown in equation (\ref{eq:half_integral}). \hfill$\Box$

\begin{figure}[]
  \centering
  \includegraphics[width=0.4\linewidth]{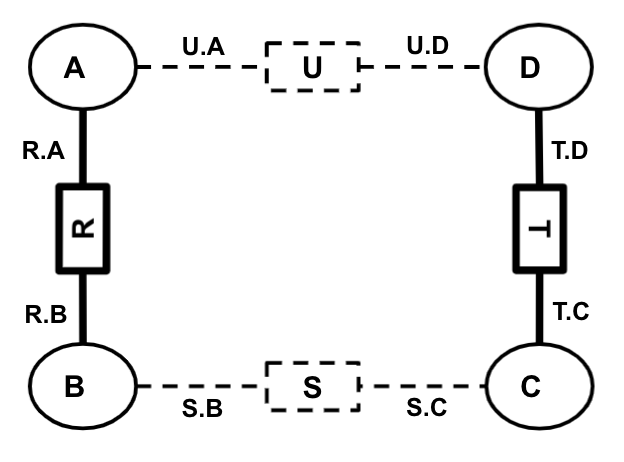}
  \caption{TAG plan of a 4-way cycle query in Example~\ref{ex:ir_cost}.}
  \label{four_cycle}
\end{figure}
\begin{newexample}
\label{ex:ir_cost}
Consider a $4$-way cycle query $R(A,B) \Join S(B,C) \Join T(C,D) \Join U(D,A)$ in Figure~\ref{four_cycle}. A fractional edge cover solution of the query is $(1,0,1,0)$ with the fractional edge cover number that is equal to $2$, and so in the case of the worst-case instance the maximum possible output size is $O(IN^2)$. Let the weights of relations $R$ and $T$ to be equal to $1$ (shown as solid lines), and the weights of $S$ and $U$ to be equal to $0$ (shown as dashed lines). Relations $R$ and $T$ form a union of stars, which we can compute by applying the Cartesian product algorithm. Lets store the result of the Cartesian product at $R$-tuple vertices. Then apply the acyclic join algorithm to join the Cartesian product result with relations $S$ and $U$. Observe that we only need to semi-join relations $S$ and $U$ with the Cartesian product result. Thus, running a bottom-up reduction phase is sufficient. 
The overall cost of evaluating a $4$-way cycle query is dominated by the cost of the Cartesian product, $O(IN^2)$, which does not exceed the worst-case output of the query.
\end{newexample}
\textbf{Extending half-integrality lemma.} Our TAG model allows us to apply the half-integrality property to decompose a query into a collection of cycles and/or union of stars on queries, where relations have more than two attributes, as well as on queries, where join conditions are not necessarily on a single attribute.
Thus, we generalize the above result (Lemma~\ref{lem:ir_cost} ) from queries on binary relations to arbitrary queries. 

Recall that in TAG model each attribute vertex is connected to its corresponding tuple vertex.
We reduce a TAG instance of a multi-attribute relation to a TAG instance of a binary relation, by using only two attributes vertices (e.g. attributes that are used in join conditions) of each tuple vertex.
Furthermore, in Section~\ref{sec:tag_plan} we also note that for join queries it is sufficient to only create TAG plan nodes for the join attributes.
Therefore, by solving a query on binary relations we solve a query on multi-attribute relations.
The values of all attributes are obtained from the tuple vertices during the collection phase, as shown in lines 32-36 of Algorithm~\ref{alg:acyclic_join}.
Alternatively, an attribute vertex can retrieve the values of the rest of the attributes of the tuple it belongs to with an extra communication round, whose cost is bounded by the size of the input.
\begin{figure}[]
  \centering
  \includegraphics[width=0.5\linewidth]{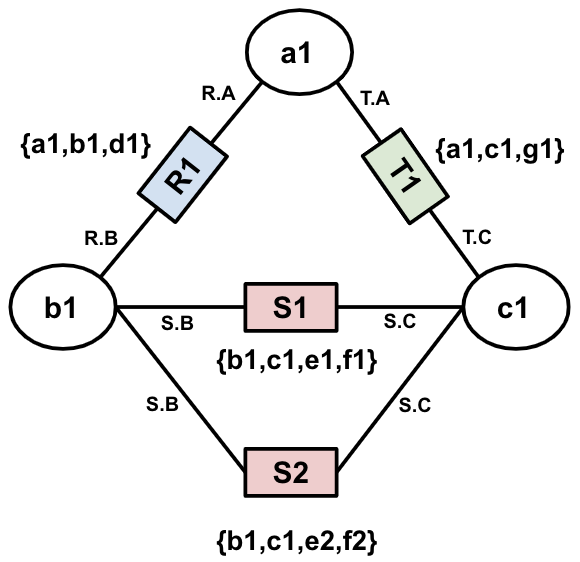}
  \caption[]{TAG instance for a triangle query with multi-attribute relations.}
  \label{triangle_multi}
\end{figure}
Figure~\ref{triangle_multi} shows the example TAG instance for a triangle query, where relations have more than two attributes:
\begin{displaymath}
R(A,B,D) \Join S(B,C,E,F) \Join T(C,A,G).
\end{displaymath}
The vertices corresponding to non-join attributes (e.g. attributes $E$ and $F$ of relation $S$) are not shown, since their values are also stored at the tuple vertices they belong to.


Join conditions on multiple attributes are handled as described in Section~\ref{sec:multi_attribute_join}, where we reduce a multi-attribute join condition to a join on a single attribute by adjusting the reduction phase.

\subsubsection{Step (2): Compute Acyclic Join}
\begin{figure}[t!]
  \centering
  \includegraphics[width=\linewidth]{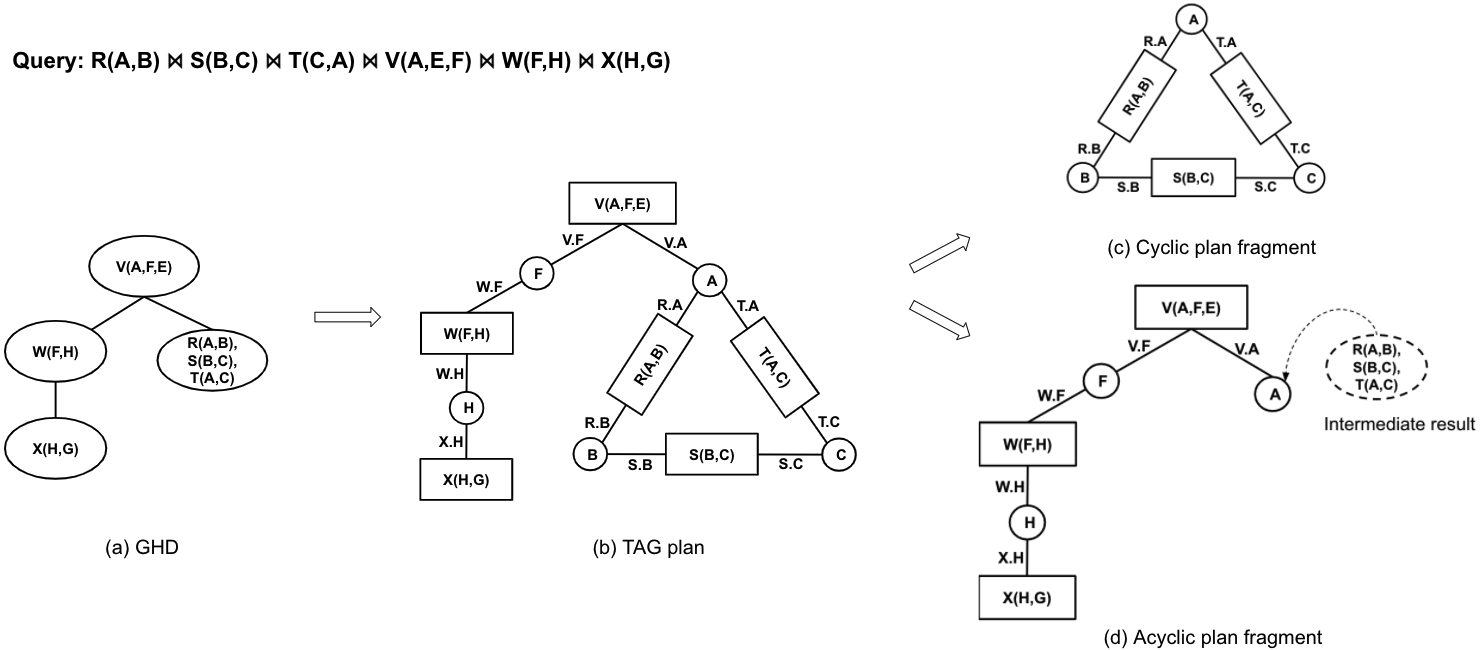}
  \caption{TAG plan of the query in Example~\ref{ex:arbitrary_query}}
  \label{query_plan_2}
\end{figure}

When computation of subqueries  is done, we continue with the acyclic TAG-plan fragment, that includes vertices where intermediate results are stored.
Run the multi-way acyclic join algorithm as described in Section~\ref{sec:acyclic_multiway} (Algorithm~\ref{alg:acyclic_join}) to get the final join result.
Lets illustrate an example execution of TAG-join algorithm on a query.
\begin{newexample}\label{ex:arbitrary_query}
Figure~\ref{query_plan_2} shows an example query, and it's corresponding tree decomposition in Figure~\ref{query_plan_2}(a).
Following the steps in Section~\ref{sec:tag_plan} we construct a TAG plan in Figure~\ref{query_plan_2}(b).
Since one of the bags of a GHD contains three relations $R, S$ and $T$, we repeat the same steps for each relation in the bag.
We start with evaluating a subquery, corresponding to the cyclic fragment of the plan, specifically the triangle query in Figure~\ref{query_plan_2}(c).
Activate $C$-attribute vertices and apply the procedure as described in Section~\ref{alg:wcoj-triangle}, i.e. propagate the values via paths that lead to $A$-attribute vertices.
Note that we can reduce the input more aggressively if $A$-attribute vertices check whether they have V.A edge before intersecting received values.
This way, we avoid computing triangles for $A$-attribute vertices that do not even connect with the rest of the query.
The triangle query results are computed and stored at $A$-attribute vertices, i.e. the intermediate results are distributed among all $A$-attribute vertices that are part of the triangle. 

Once done with the triangle, continue with the computation of the acyclic join plan in Figure~\ref{query_plan_2}(d) following Section~\ref{sec:acyclic_multiway}. Using Algorithm~\ref{alg:vc_program} generate a list of labels to drive the vertex program described in Algorithm~\ref{alg:acyclic_join}.
The computation completes at $V$-tuple vertices, which then can output the result.
\end{newexample}

\subsubsection{Cost Analysis}
For a given input query $q$ the complexity of TAG-join algorithm is dominated by the cost of computing the subqueries corresponding to the bags of a tree decomposition (GHD).
The cost of computing each intermediate result is defined using a fractional edge cover number ($\rho^*$) of each bag $q_i$, recall AGM bound \cite{agm}. Note that in the worst-case scenario for a given bag the result is the Cartesian product of relations that are assigned to it.
Given a tree decomposition of $q$ with a fractional hypertree width $w$, such that $\rho^*(q_i) \leq w$ and $w$ is the minimum of the widths of all tree decompositions of the query, it follows from Lemma~\ref{lem:ir_cost} that the cost of computing all intermediate results is at most $O(IN^w)$.

Then applying the acyclic join algorithm we compute the final output, $OUT$.
The acyclic join algorithm incurs the communication and computation cost that is linear in the size of the input and the output (see Section~\ref{sec:acyclic_multiway}). Lets denote the input size of the query with computed intermediate results as $IN'$, then $IN' \leq IN^w$.  Therefore, the overall complexity of TAG-join algorithm is $O(IN^w+OUT)$, as claimed in Theorem~\ref{th:main}.
\begin{newexample}
Continuing with the query from Example~\ref{ex:arbitrary_query}, we now estimate the total cost of the algorithm.
Figure~\ref{query_plan_2}(a) shows a GHD with width $w=\frac{3}{2}$.
The first step is to compute intermediate results corresponding to the triangle query, which takes $O(IN^{\frac{3}{2}})$ (see Section~\ref{sec:triangle}). And the final step of computing the acyclic plan fragment in Figure~\ref{query_plan_2}(d), costs $O(IN^{\frac{3}{2}} + OUT)$ communication and computation. Note that the $OUT$ itself is bounded by $O(IN^3)$, which is the worst-case output size of the given query (AGM bound).
\end{newexample}

\subsubsection{Comparison to Other Algorithms} 
Our TAG-join algorithm is broadly inspired by GYM \cite{afrati_3,koutris_book}, the generalized version of parallel Yannakakis for arbitrary join queries. GYM employs generalized hypertree decompositions as input logical plan, and starts by computing each bag of a GHD to then apply Yannakakis reduction on the resulting intermediate results. 
The same GYM-style approach is also used in the EmptyHeaded engine \cite{eh}, and in the InsideOut algorithm \cite{faq}. Both use worst-case optimal algorithms \cite{nprr,leapfrog,skew_strikes} to compute each bag of a GHD, and achieve the same total complexity as TAG-join. However, recall that worst-case optimal algorithms heavily rely on organizing the input into index structures based on a global attribute order, which are expensive to compute and maintain.

\section{Beyond Equi-Joins}\label{sec:beyond-joins}
TAG-join is
compatible with the efficient evaluation of other algebraic operations.
Small edits to the vertex program (Algorithm~\ref{alg:acyclic_join}) allow the seamless interleaving of algebraic operations,
supporting classical optimizations such as pushing selections, 
projections and aggregations before the join.
We also show how we deal
with aggregations, outer joins and subqueries, including correlated subqueries.
\paragraph{Selections}
\revise{Pushing selections before joins translates in our setting to vertices checking the
selection condition as early as possible.
Conditions involving a single attribute are checked in parallel 
by the corresponding attribute vertices during the reduction phase (by adding the selection 
to line 12 in Algorithm~\ref{alg:acyclic_join}). Attribute vertices that fail the selection deactivate themselves, reducing overall computation and communication.
Conditions involving multiple attributes are also applied in parallel by attribute vertices
but need to wait for the earliest round of the collection phase where collected intermediate tuples contain the relevant attributes.
This is achieved by adding the selection to line 31 in Algorithm~\ref{alg:acyclic_join},
together with a check of the current label (set in line 21) to identify the round.
}
\paragraph{Projections}
\revise{Pushing projections early is beneficial for the collection phase.
Although it does not reduce the number of sent messages, it affects their size by reducing the number of attributes of the comprised tuples. Projection pushing is implemented by application to the local joins in lines 34 and 36 of Algorithm~\ref{alg:acyclic_join}. Which columns
can be projected away depends on the round, which in turn is given by the current label (set in line 21).
}
\paragraph{Aggregations}
\revise{
The aggregation scheme is inspired
by aggregation over hypertree decompositions in factorized databases \cite{fdb_1,fdb_2}, since our
TAG plan~\footnote{Note that the TAG plan should include a set $G$ of attributes from a GROUP BY clause, such that an attribute is a root node or a child of another attribute in $G$ \cite{fdb_1}.} is based on a GHD (recall Section~\ref{sec:tag_plan}).
The scheme includes the classic optimization technique of pushing grouping and aggregation before the join \cite{eager_agg}. 
Aggregates are computed during the collection phase as soon as the intermediate tuples contain the relevant
attributes, by using a modification to line 31 in Algorithm~\ref{alg:acyclic_join} 
and by checking the current label to identify the aggregation rounds.}

We distinguish three types of aggregation:
\begin{itemize}
    \item Local aggregation (LA) corresponds to SQL queries with GROUP BY on one attribute, or multiple attributes where one attribute functionally determines the others. 
    \item Scalar aggregation computes a single tuple of scalar values. 
    The aggregation is computed in parallel bottom-up, and once the root is reached, all active vertices need to send their computed aggregate to a global 'aggregation' vertex whose id is known to all to get a final result. 
    \item Global aggregation (GA) uses a multi-attribute GROUP BY clause, such that the attributes do not determine each other. This also requires vertices to communicate with a global 'aggregation' vertex to output a final aggregation.  
\end{itemize}
\begin{figure}[]
  \centering
  \includegraphics[width=0.8\linewidth]{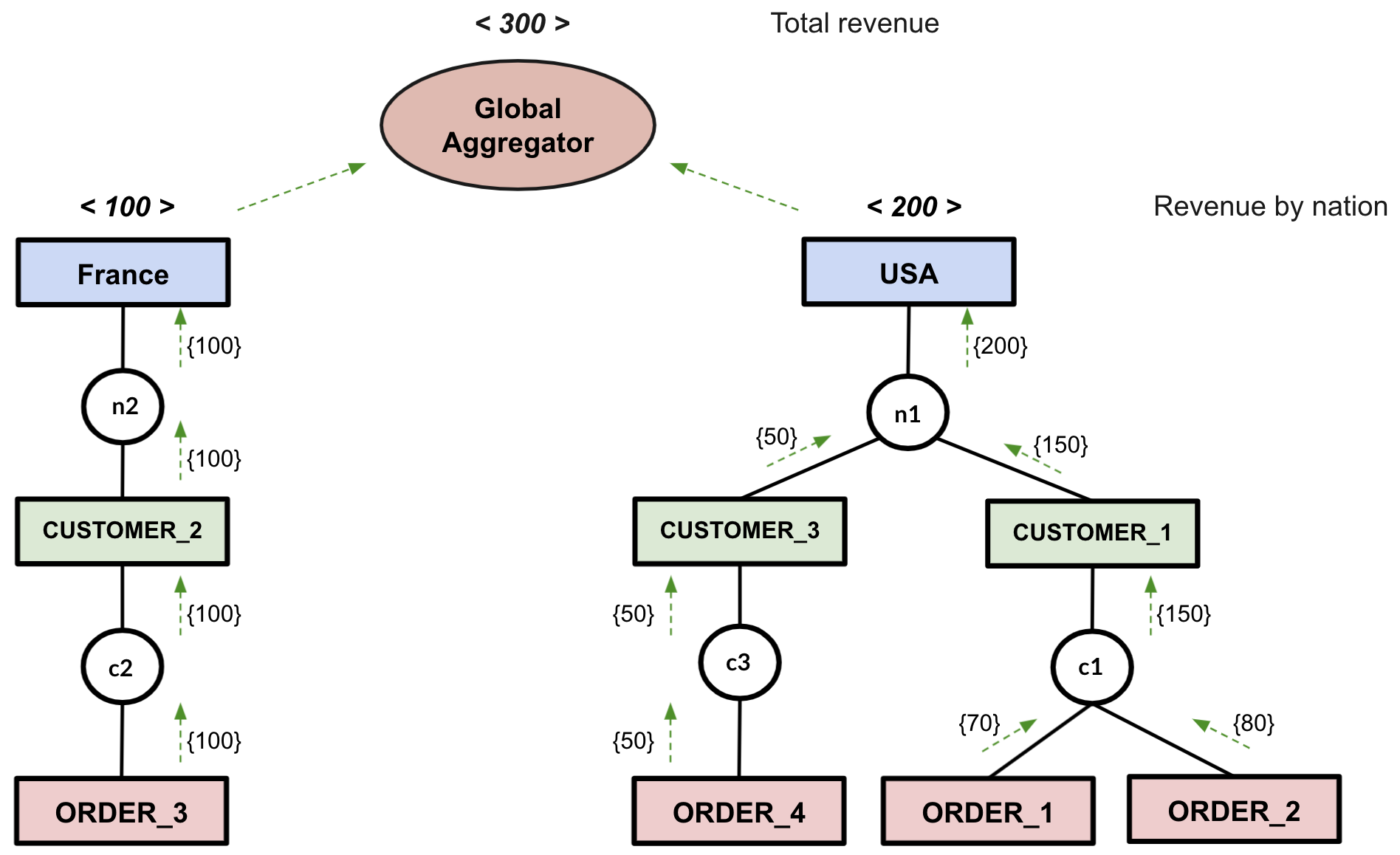}
  \caption{Example of aggregations: local aggregation to compute revenue by nation and global aggregation to compute the total revenue.}
  \label{aggregation}
\end{figure}
Local aggregation benefits the most from vertex-centric computation: 
aggregation within each group is computable
by the attribute vertex representing the group key, in parallel to the other groups.
While global aggregation is bottlenecked by a single global aggregator vertex that receives messages from all vertices.
The example of local and global aggregations is shown in Figure~\ref{aggregation}. Local aggregation is used to compute revenue by nation. Each group by key (i.e. nation name) corresponds to a vertex, thus aggregation for each group can be computed independently of the other groups. And all groups are done in parallel. The global aggregator vertex is used to compute the total revenue over all nations, i.e. scalar global aggregation. Note that, instead of sending values of $ORDER$ tuples directly to the global aggregator vertex, we can compute local aggregates for each nation in parallel, and then send the results of partial aggregation. This will not remove the bottleneck of a single global aggregator vertex, but can greatly reduce the number of messages it needs to process, since there are a lot less $NATION$ vertices than there are $ORDER$ vertices.

We employ eager aggregation \cite{eager_agg}, an optimization technique of pushing down the group by, i.e. perform early group by prior to the join. This optimization technique is justified when filtering and/or join conditions are not as selective, hence early group by helps to reduce the cost of join (computation and communication cost).
Eager group by reduces the number of supersteps of the computation since there is no need to go top-down in a reduction phase to mark join edges and then in a bottom-up traversal again in order to collect values that need to be aggregated, and as a result the number of messages that are sent is also reduced. Instead in one bottom-up reduction phase the values are aggregated and join conditions are checked, i.e. interleaving join and group by operators. Modifications to the vertex program in Algorithm~\ref{alg:acyclic_join} involve line 9, where the received values in the incoming queue are combined (e.g. sum), and line 13, where the resulting aggregate value, or in case of a first superstep a value from a tuple to be aggregated is sent as a message further.

\paragraph{Outer Joins}
An outer join is a variation of a join operator, where the output can contain dangling tuples, i.e. tuples that do not match with any tuple of the other relation in the join attributes. Most commercial relational database systems do provide an outer join operator.
The are three types of an outer join, depending on tuples of which side of a join are added to the output regardless of whether they have a matching tuple: left, right and full (both sides).

With simple adjustments, outer joins can be computed by a
TAG-join algorithm. Consider a two-way join example from Section \ref{sec:single_attrib_join}, let $R$ be the left side relation and $S$ be the right side relation. To compute left outer join each join $B$-attribute vertex needs to have at least one outgoing edge with label $R.B$ connecting it to the left side relation in order to stay active and to continue the computation. Qualifying $B$-attribute vertices send messages to tuple vertices along edges with label $R.B$ and possibly along $S.B$ edges. Then, we start the collection phase, where activated tuple vertices send back their values. If $B$-attribute vertex is not connected to any $S$-tuple vertex, then the output tuples are constructed from the received $R$-tuples only with missing values (e.g. $NULL$) of the right relation $S$.
For a right outer join, $B$-attribute vertex must have at least one edge with label $S.B$ to participate in the computation. Therefore, in the collection phase each $B$-attribute vertex receives at least one $S$-tuple that is part of the output, while $R$-tuple values can be missing.
To compute a full outer join we skip the reduction phase, since dangling tuples from both relations are allowed. We start the computation and go directly into the collection phase, where activated tuple vertices of both relations $R$ and $S$ send their values to $B$-attribute vertices, which then construct the output tuples. 

\paragraph{Subqueries}
Subqueries are an important and powerful component of SQL. Relational database benchmarks such as TPC-H \cite{tpch} and TPC-DS \cite{tpcds} include a lot of queries that use subqueries in various SQL clauses: $SELECT$, $FROM$, $WHERE$, $HAVING$ and $WITH$. A SQL query can contain multiple levels of nested subqueries. Subqueries are used to return either scalar values or a relation (i.e. multiple tuples/rows). In a vertex-centric approach subquery results are computed in parallel and stored at vertices in a distributed way. In case of scalar values, we use a global aggregator vertex to compute and store the scalar result. 

To evaluate subqueries we rely on navigational strategies with forward lookup and reverse lookup \cite{subquery_exe}  used in traditional RDBMSs. 
With reverse navigational lookup we start the traversal of a TAG plan from a subquery (i.e. relations used in a subquery), and then navigate towards an outer query. It especially applies to subqueries that appear in $FROM$ and $WITH$ clauses, where subquery results are used as inputs to compute the outer query.
On queries, where a subquery is related to an outer query, we apply a forward lookup such that a subquery only processes vertices that are relevant to the outer query (see correlated subqueries).

\paragraph{Correlated Subqueries.} A correlated subquery is a subquery that depends on the values of the outer query. We employ a navigational execution strategy with a forward lookup \cite{subquery_exe}, i.e. start with an outer query, and in a tuple-at-time fashion invoke a subquery.
Our vertex-centric approach 
naturally supports parallelization: tuple vertices matching the outer query are activated in parallel and their subquery calls execute in parallel.

\paragraph{Semi-join and Anti-join.}
Recall that semi-join $R(A,B) \lJoin S(B,C)$, retrieves precisely those $R$-tuples that join with at least one $S$-tuple.
Therefore, to perform a semi-join 
we activate $R$-tuple vertices, which then send messages to their corresponding $B$-attribute vertices. Each $B$-attribute vertex checks its outgoing edges, and sends message back to $R$-tuple vertices only if it has at least one outgoing edge with label $S.B$ (i.e. connected to $S$-tuple vertex). After this superstep, only $R$-tuple vertices corresponding to the semi-join result are active and can output their values.

Anti-join performs the opposite of a semi-join operator, and returns $R$-tuples that do not match any $S$-tuple in join attribute. To obtain the anti-join result, we modify the vertex-centric semi-join computation, such that each $B$-attribute vertex sends message back to $R$-tuple vertices only if none of its outgoing edges have label $S.B$.

We apply semi-join and anti-join strategies when evaluating subqueries with $IN$, $EXISTS$, $NOT$ $IN$ or $NOT$ $EXISTS$ constructs.

\begin{figure}[]
  \centering
  \includegraphics[width=\linewidth]{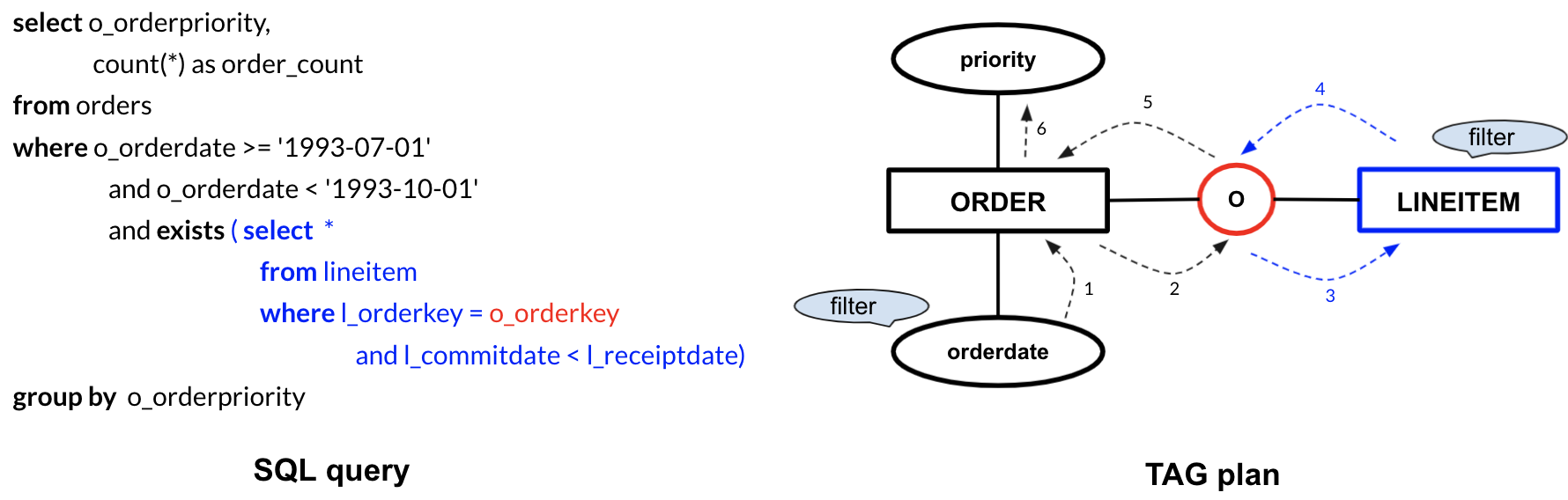}
  \caption{Correlated subquery example.}
  \label{corr_subquery}
\end{figure}
\begin{newexample}
Lets consider SQL query from the TPC-H benchmark, shown in Figure~\ref{corr_subquery}.
It contains a correlated subquery using $EXISTS$ construct. The TAG plan of the query shows which part of the plan corresponds to the subquery (in blue), and the attribute vertex $O$ (in red) defines the correlation, i.e. $orderkey$ attribute that relates the subquery to the outer query.
Edge labels are not shown to avoid clutter.
The dashed arrows indicate the order of the traversal steps of the vertex-centric computation, while the blue bubble indicates a filter condition to be checked by a vertex.  
We start the evaluation of the query from the date attribute vertices and apply the filtering condition such that only the vertices whose date values are in the given range send messages to $ORDER$ tuple vertices.
From $ORDER$ tuples we navigate to $orderkey$ attribute vertices, depicted as $O$, i.e. tuple vertices of $ORDER$ send messages to attribute vertices, and activate them for the next superstep of the computation.
We then continue to evaluate the correlated subquery, where from the orders ($O$ attribute vertices) that are discovered by the outer query we traverse to $LINEITEM$ tuple vertices.
Tuple vertices, that satisfy the filtering condition represent the result of the subquery. Notice that, we essentially perform the semi-join and evaluate $EXISTS$ operator by navigating from filtered $LINEITEM$ tuples back to $O$ attribute vertices. In turn, activated $O$ attribute vertices send messages to $ORDER$ tuple vertices.
This completes the reduction phase, and we now have active tuple vertices of $ORDER$ relation that contribute to the output.
The collection phase traverses from $ORDER$ tuple vertices to $priority$ attribute vertices. Each $ORDER$ tuple vertex sends a message containing value $1$. Attribute vertices aggregate the received messages by summing the values to compute the count of orders. Thus, we obtain the final result of the order counts for each order priority. The result is distributed across attribute vertices corresponding to the $priority$ attribute values.
\end{newexample}

\begin{figure*}[t]
  \centering
  \includegraphics[width=\linewidth]{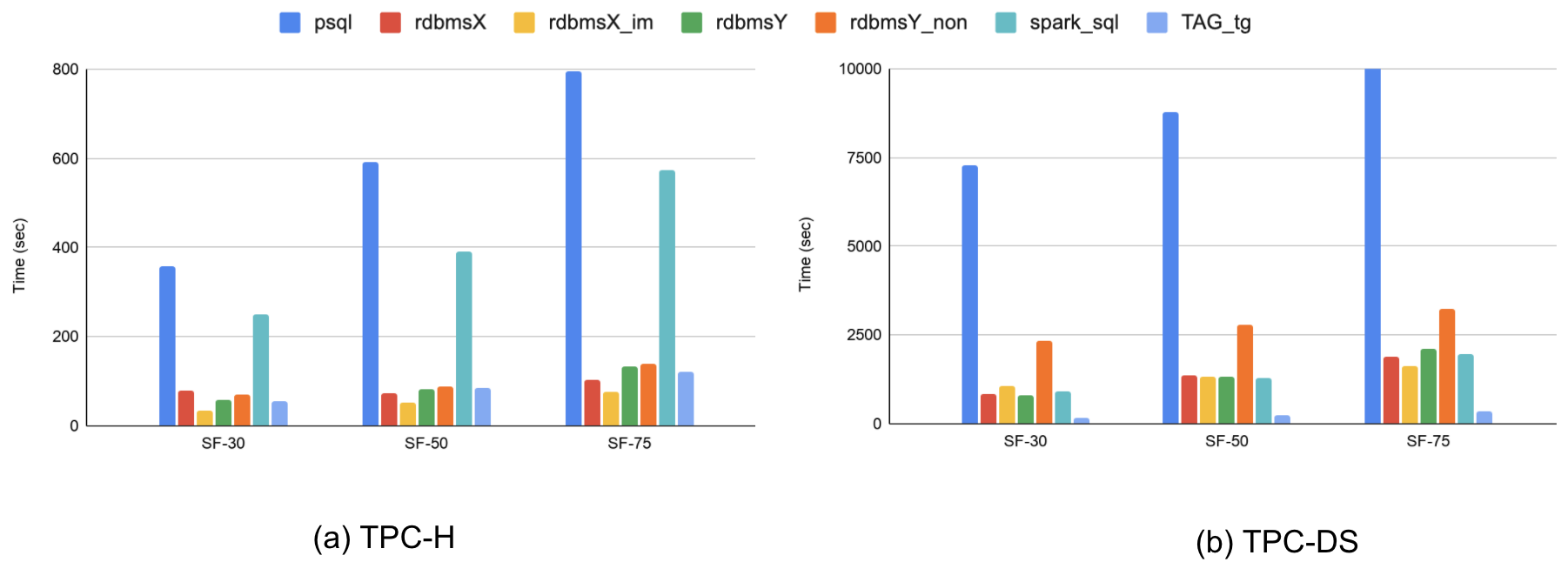}
  \caption{Aggregate runtimes of TPC-H and TPC-DS queries over datasets of scale factors 30, 50 and 75.
  }
  \label{fig:total_runtimes}
  \Description{}
\end{figure*}
\section{Experiments}\label{sec:experiments}
\revise{Our vertex-centric SQL evaluation scheme applies to both intra-server thread-based and to distributed cluster parallelism.
The bulk of our experiments (Sections ~\ref{sec:exp-data-loading},~\ref{sec:centralized-tpc-h},
~\ref{sec:centralized-tpc-ds}, ~\ref{sec:mem_usage})
evaluates how our approach enables thread parallelism}
in the {\em comfort zone} of high-end RDBMSs: running the benchmarks they are traditionally
tuned for, on a multi-threaded server with large RAM and SSD memory holding 
all working set data in warm runs. 
\revise{
We also carry out preliminary experiments evaluating the ability to exploit
parallelism in a distributed cluster, where we compare our approach against
Spark SQL (Section~\ref{sec:distr_exp}).
}
We detail our performance comparisons below but first we summarize the experimental results.

Figure~\ref{fig:total_runtimes} shows the aggregate runtimes (i.e. summed over all queries) 
for the TPC-H and TPC-DS query workloads in \revise{single-server mode}. 
For each benchmark we performed three sets of experiments with varying data sizes. 
In aggregate, the TAG-join approach outperforms \revise{all} relational systems (5x-30x speedup) on TPC-DS queries. 
On TPC-H queries, it is much faster than PostgreSQL \revise{and Spark SQL}
and competitive with all others except for
\dbo IM \revise{(whose speedup does not exceed 1.6x)}.
As the drill-down into our measurements shows, the TAG-join excels particularly
at computing local-aggregation queries and, regardless of aggregation style,
PK-FK join queries and queries with selective joins, 
\revise{outperforming even \dbo IM on these query classes}.

\revise{
Figure~\ref{fig:distr_summary} shows that our approach outperforms
Spark SQL in both aggregate runtime and network
communication in our distributed experiments.}

\subsection{\textcolor{black}{Single-
Server} Experiment Setup}
\subsubsection{Datasets and Queries}

We evaluate our approach using two standard relational
database benchmarks TPC-H \cite{tpch} and TPC-DS \cite{tpcds}.
The schema of TPC-H is defined as a pure 3rd Normal Form (3NF) schema and comprised of 8 separate tables with low number of columns. TPC-DS implements a multiple snowflake schema with 7 fact tables and 17 dimension tables, where tables are much wider compared to TPC-H, with an average of 18 columns \cite{tpcds_1}. Both benchmarks provide a tool to generate a dataset of a certain size specified by scale factor (SF), e.g. SF = 1 generates a dataset of size 1GB. TPC-H uses a purely synthetic data generator and all tables scale linearly with the database size (SF). TPC-DS employs a hybrid approach of data (number of tuples) and domain (value set) scaling based both on synthetic and real world data distributions, and as a result provides a more skewed dataset. In TPC-DS while fact tables scale linearly, dimension tables scale sub-linearly as the scale factor of the database grows. Moreover, in TPC-DS any column except for primary keys can have missing (NULL) values. 

The query workload of the TPC-H benchmark consists of 22 queries: 
20 among them with 2 to 8 joins, the remaining 2 queries are single table scans. 
All but 1 are acyclic, the exception is a five-way cycle query. The TPC-DS benchmark offers a more realistic and challenging query workload that contains 99 acyclic queries in total \cite{tpcds_2}, of which we evaluate 84 queries in our experiments. We discarded 15 queries that contain functions that are not supported by the vertex-centric platform we used
(and are not in the scope of this work), such as ranking function (rank over()), 
extracting a sub-string, and computing standard deviation. 
We also exclude a scenario containing iterative queries. 
The 84 TPC-DS queries in our experiments feature between 3 to 12 joins each, 
including joining multiple fact tables and multiple dimension tables, 
and joins between large dimension tables
(so not all joins are PK-FK, e.g. q54).
The queries are of varying complexities, containing aggregation and grouping and 
correlated subqueries. All queries are run without the ORDER BY and LIMIT clauses as 
we left top-k processing outside the scope of this paper.

\subsubsection{The Vertex-Centric System We Used}\label{sec:vc_used}
We implemented our approach on top of TigerGraph~\cite{TG19}, a graph database 
system that supports native graph storage and a vertex-centric BSP computational model
\revise{that takes advantage of both thread and distributed cluster parallelism}. 
TigerGraph offers a graph query language that allows developers to 
express vertex-centric programs concisely. The mapping from the query
to the vertex program is straightforward, well-defined~\cite{TG19,TG20}, 
and is not where the optimizations kick in (we checked with the TigerGraph engineers). 
\revise{This ensures that the queries we wrote execute our intended vertex-centric programs
faithfully, without compiling/optimizing them away.}

\revise{Some systems offer vertex-centric APIs which are implemented internally via relational-style joins (e.g. \cite{eh,pregelix}), thus being unsuited for our experiment. In contrast, TigerGraph features native implementation for the vertex-centric primitives we describe here. Inspection of the queries we published with the supplemental materials reveals that they each are expressed as a sequence of separate one-edge hops, i.e. vertex-centric steps where edge sources send messages to edge targets.}
All queries are published in the additional submission materials~\cite{extended-version}.

We used the free TigerGraph 3.0 Enterprise Edition in experiments
\revise{and ran it in main-memory mode}.
We denote our implementation with \textbf{TAG\_tg} in figures and tables below.

\subsubsection{The Comparison Systems}
We compared our TAG-join implementation against popular relational database systems:\\

\emph{PostgreSQL 12.3} is a popular open-source relational database implemented as a row store (\textbf{psql} in tables).
    
\emph{\dbo} is a commercial relational database with row store support (release version from 2018). \dbo offers an In-Memory (IM) feature that uses an in-memory column store format. This format accelerates query processing by enabling faster data scans, aggregation and joins. 
We ran the queries with two settings: traditional row store format (\textbf{\dboSmall}) and 
dual format with in-memory column store enabled (\textbf{\dboSmallIM}).


\emph{\dbm} is a free developer edition of a well known commercial database with row store support (release version from 2019). We run \dbm with two settings: clustered (\textbf{\dbmd}) and non-clustered primary key (\textbf{\dbmnon}).

\revise{
\emph{Spark SQL 3.0.1}~\cite{spark_sql} is a module of Apache Spark~\cite{spark}
supporting relational (and JSON) data processing. For the purpose of our experiments, 
we include it in the collective term "relational engines" (\textbf{spark\_sql}).
Spark is one of the most widely-used distributed general purpose cluster-computing engine, thus we evaluate it against our approach in the single-server setting as well as the cluster setting (see Section~\ref{sec:distr_exp}).\\
}

For all relational database systems we tuned memory parameters (e.g. buffer cache size) to ensure that the entire database as well as intermediate query results can fit into memory
(we monitored disk usage during the experiments, confirming that all RDBMS warm
runs performed no disk access).
For PostgreSQL, we adjust the following key parameters: $shared\_buffers$
, $work\_mem$
, and $temp\_buffers$. 
\dbm uses automatic memory management, and by default acquires as much as possible memory. We observed that \dbm allocated enough buffer space to fit the entire input database. For \dbo we enable automatic memory management as well.

We enabled parallel execution of queries for all of the RDBMSs without forcing a specific degree of parallelism, letting their optimizer decide on how many workers to use during execution. 
For PostgreSQL we tuned the parallel query specific parameters (e.g $max\_parallel\_workers$, $max\_worker\_processes$ and $max\_parallel$ $\_workers\_per\_gather$) to adjust them to the number of available threads on the machine. These parameters define the maximum number of workers that the query optimizer will consider when planning the query. 
For commercial databases we enable the query optimizer to automatically determine whether to execute a query in parallel or not, and what degree of parallelism (similar to the number of workers in PostgreSQL) to use by setting their corresponding parameters.

With in-memory column store enabled \dbo populates the entire database into the column store at the default memory compression level, which is optimized for query performance as opposed to space economy since there is enough storage space for both of the datasets. 
Once all tables are populated into the In-Memory column store, in-memory storage indexes are created automatically on each column. 

\revise{For Spark, we use Parquet \cite{parquet}, a compressed columnar file format, as a data source, for which it supports column pruning and pushing down filter predicates. We also enable in-memory caching of the input data in our experiments.}

\subsubsection{Hardware}
For all systems we used an AWS EC2 r4.8xlarge instance. This instance type has 2.3 GHz Intel Xeon E5-2686 v4 processor with 32 vCPU count (number of supported parallel threads), 
244 GB of memory and 500GB SSD drive, running Ubuntu 16.04. 
For \dbo we used Linux 7.6.

\subsubsection{Methodology} 
The queries are evaluated on datasets obtained with the benchmark generators,
at scale factors 30 (30GB), 50 (50GB) and 75 (75GB).
We stopped at SF-75 to make sure that the input database can be cached without approaching the main memory limit of the machine.
Each dataset is supplied with primary and foreign key indexes. Each query is executed 11 times, the first run to warm up the cache and the remaining 10 runs to compute the average runtime. We impose a timeout of 30 minutes per execution. 

\begin{figure*}[t]
  \centering
  \includegraphics[width=\linewidth]{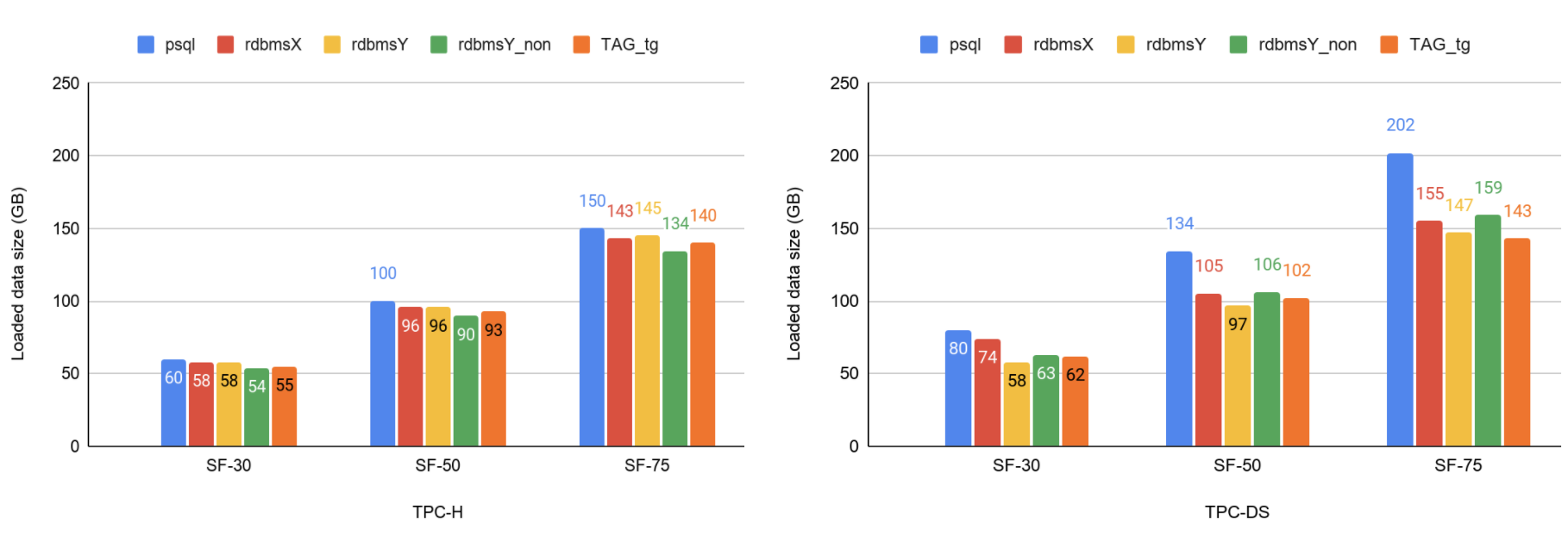}
  \caption{The sizes of the loaded datasets across different scale factors. The numbers on top of the bars correspond to the total data size, including the sizes of primary and foreign key indexes.}
  \Description{}
  \label{fig:load_size}
\end{figure*}
\subsection{\textcolor{black}{Single-Server} Data Loading Results}\label{sec:exp-data-loading} 
\begin{table}[t]
\caption{Loading times for TPC-H dataset including PK-FK index creation time, shown in seconds.}
\begin{tabular}{l|ccc}
\toprule
           & \multicolumn{3}{c}{TPC-H}                                                         \\
           & \multicolumn{1}{l}{SF-30} & \multicolumn{1}{l}{SF-50} & \multicolumn{1}{l}{SF-75} \\
\midrule
psql      & 2085.21                  & 3590.27                   & 5401.81                   \\
\dboSmall     & 2007.04                   & 3774.92                   & 7074.54                  \\
\dbmd      & 3123.52                   & 5343.32                   & 6963.37                   \\
\dbmnon & 2619.53                   & 4405.66                   & 6582.68                   \\
TAG\_tg    & 2942.47                   & 5146.17                   & 7982.37                  
\end{tabular}
\label{tab:load_timeTPCH}
\end{table}

\begin{table}[t]
\caption{Loading times for TPC-DS dataset including PK-FK index creation time, shown in seconds.}
\begin{tabular}{l|ccc}
\toprule
           & \multicolumn{3}{c}{TPC-DS}                                                        \\
           & \multicolumn{1}{l}{SF-30} & \multicolumn{1}{l}{SF-50} & \multicolumn{1}{l}{SF-75} \\
\midrule
psql      & 3419.76                   & 6024.05                   & 9622.38                   \\
\dboSmall     & 6093.7                    & 10196.67                  & 15473.27                  \\
\dbmd      & 2874.4                    & 5063.6                    & 7595.4                    \\
\dbmnon & 3475.41                   & 5815.81                   & 8723.72                   \\
TAG\_tg    & 3107.26                   & 5254.06                   & 8019.52                  
\end{tabular}
\label{tab:load_timeTPCDS}
\end{table}

We measured both loading time (including index creation time for
the RDBMSs) and loaded data size (including index size for RDBMSs).
Recall that no indexes are constructed for the TAG representation
since the attribute vertices act as indexes implicitly.

\revise{Spark SQL does not perform the heavy-weight storage management of full-blown database management systems (including TigerGraph), and thus avoids their overhead. It's mainly a data processing engine, that accesses data from either local files, distributed files systems (HDFS, S3) or other databases. In order to avoid apples-to-oranges comparison we do not cover Spark SQL in this section.
}

The loading times are shown in Table~\ref{tab:load_timeTPCH} and Table~\ref{tab:load_timeTPCDS}, corresponding to TPC-H and TPC-DS datasets respectively.
With the exception of \dbo on TPC-DS dataset, the total loading times are roughly comparable across all systems for different scale factors.

Figure~\ref{fig:load_size} depicts the sizes of the loaded data for all the considered systems and scale factors.
The datasets are generated by the tools provided with the benchmark suites, and are loaded into databases using their respective commands for bulk data load. For relational systems, primary and foreign key indexes are built on the loaded data, as prescribed by the TPC benchmark protocol. 
All RDBMSs we deployed organize indexes in a B-tree structure by default.

For $TAG$ graph we materialized all of the the attribute vertices of integer type, that essentially correspond to PK and FK indexes created for relational systems, date and string types (e.g. attributes used in $GROUP$ $BY$ clause and filtering).
However, we do not load float values and some string attributes (e.g. string attributes corresponding to comments or long descriptions), as attribute vertices. These attributes are not used as join conditions in the given query workload, and it is sufficient to store these values in their respecitve tuple vertices.
We do create definitions of all vertex (tuple and attribute) and edge types, and more attribute vertices can easily be materialized as needed without the need to reorganize the existing graph.  

With in-memory feature enabled \dbo allows to store data in compressed columnar format. We use the default compression method to get the best query performance as recommended by the documentation. Table~\ref{tab:load_im} shows the sizes of the in-memory store after data is transformed into columnar format, as well as the original data size, excluding the index sizes. This is not the size of the area that is allocated for in-memory in order to enable the feature, but the size of the in-memory segments that are actually populated with data.

\revise{
The gist of the experiment is that we observed similar loaded data size for
all systems (within 10\% of each other except for the more wasteful PostgreSQL) 
and similar loading times (also within roughly 10\% of each other,
except for \dbo IM which takes double the time of the others on TPC-DS).
This testifies to the absence of time and space overhead for loading data 
as a graph vs loading it into an RDBMS. 
}

\subsection{\textcolor{black}{Single-Server} TPC-H Results}\label{sec:centralized-tpc-h}
Figure~\ref{fig:total_runtimes}(a) shows the aggregate run times of TPC-H queries over datasets of different scale factors.
\revise{All 22 TPC-H queries contain a certain type of aggregation as discussed in Section~\ref{sec:beyond-joins}, and TAG-join performs especially great on queries with local aggregation and on queries that contain a correlated subqueries.

PostgreSQL, \dbm and \dbo support three main join algorithms, i.e. nested loop, hash join and sort-merge, and a query optimizer of each database chooses one of them to evaluate a given query. Aggregation with GROUP BY key is implemented using either hash or sort methods.

In-memory column stores used in \dbo can drastically improve data scans by working on compressed 
columns directly, and this speeds up the application of filters such as 
$<$,$>$, $=$ and $IN$, which all of these queries do contain. 
Aggregation operators also benefit greatly from columnar layout \cite{column_db1}.  

Spark SQL enables relational processing via DataFrame API. A DataFrame is a ditributed set of rows with a schema, i.e. an equivalent of a relational table in databases.   
Similarly to RDBMSs, Spark SQL includes a rule-based and a cost query optimizer responsible for query planning. A join is evaluated using either sort-merge, shuffle hash join or broadcast join method. Spark SQL uses a compressed and partitioned columnar storage for more efficient query processing.

TigerGraph's main features are described in Section~\ref{sec:vc_used}. Aggregation is implemented via accumulators as detailed in \cite{TG20}. Accumulators are the data containers that store a value, and then aggregate inputs into it. Local aggregation is done via so called vertex accumulators, i.e. each vertex has its own local accumulator. While global aggregates are computed using global accumulators, that are accessible by all active vertices.     
}

In aggregate, TAG-join approach is 6.5x faster than PostgreSQL, \revise{4.7x faster than Spark SQL}, and shows competitive performance with \dbm (both settings) and \dbo. 
The exception is \dbo  with In-Memory column store, which 
outperforms TAG-join by 1.6x.
Full results of individual queries are shown for all three scale factors in Tables \ref{tab:tpch_75}, \ref{tab:tpch_50}, \ref{tab:tpch_30} below.

\paragraph{LA Queries.}
TAG-join performs the best on the queries that have {\em local aggregation (LA)}, 
i.e. their GROUP BY clause features either a single attribute or multiple attributes 
where one attribute functionally determines the others. This is explained by the
fact that each group can be gathered and aggregated in parallel at the attribute vertex
corresponding to the group key.
The speedups of selected TPC-H queries at SF-75 are shown in Table ~\ref{tab:tpch_speedup}. 
TAG-join is competitive with \dbo, both row store and in-memory column store 
formats, with a few exceptions such as query q3 and q4 
where in-memory column store is faster by 1.4-1.6x.
This is expected since 
in-memory column stores improve data scans and computation of aggregate values by working on compressed 
columns directly. 
TAG-join is faster than \dbm by 1.5-2.8x and 
than PostgreSQL by 4.4-9x on LA queries. \revise{It outperforms Spark SQL by 5-8.8x on these type of queries.}
Query q5 is a 5-way cycle query, where TAG-join is 10x faster than PostgreSQL, \revise{7.6x faster than Spark SQL}, 1.5-1.7x faster than \dbo, and comparable performance with \dbm non-clustered primary key. Only \dbm with clustered primary key setting outperforms by 1.6x.

\paragraph{Correlated Subqueries} TAG-join performs great on queries that contain correlated subqueries (these are LA queries). 
For example, q2, q20, q17 and q21 in Table~\ref{tab:tpch_speedup}. On individual queries TAG-join outperforms PostgreSQL by 3-200x, \revise{Spark SQL by 8-105x}, and \dbo by 1-3x. TAG-join is competitive with \dbm and \dbo in-memory on these queries, i.e. 1-1.5x speedup.    

\begin{table}[t]
\caption{The average runtimes (in seconds) of selected TCP-H queries with local aggregation (LA) and correlated subqueries (Corr) on SF-75 for TAG-join approach, and its speedups over relational engines.}
\begin{tabular}{l|ccccccc}
\toprule
SF-75 & TAG\_tg & psql & \dboSmall & \dboSmall\_im & \dbmd & \dbmd\_non & spark\_sql\\
\midrule
\textbf{LA}        & \textbf{}       &        &        &            &       &    &         \\
\texttt{q3}       & \textbf{4.28}   & 4.4x  & 1x   & 0.6x       & 1.7  & 1.8x   & \textcolor{black}{5x}   \\
\texttt{q4}       & \textbf{1.72}  & 4.7x  & 2.1x   & 0.7x      & 2.4x  & 2.8x    &  \textcolor{black}{8.8x} \\
\texttt{q5}       & \textbf{4.76}  & 10.6x  & 1.5x   & 1.7x       & 0.6x  & 1.1x  &  \textcolor{black}{7.6x}  \\
\texttt{q10}       & \textbf{4.07}  & 7.2x   & 1.5x   & 1x       & 2.1x  & 2.4x   &  \textcolor{black}{5.4x}   \\
\midrule
\textbf{Corr}        & \textbf{}       &        &        &            &       &    &         \\
\texttt{q2}       & \textbf{0.64}   & 38.1x  & 2.9x   & 1x       & 1.7x  & 1.7x  &  \textcolor{black}{28.8x}  \\
\texttt{q17}       & \textbf{0.49}  & 206.4x  & 1x   & 1.5x       & 5.7x  & 10.6x   & \textcolor{black}{105x}   \\
\texttt{q20}       & \textbf{0.94}  & 36.4x  & 1x   & 9.6x      & 1x  & 1.5x   &  \textcolor{black}{17.5x} \\
\texttt{q21}       & \textbf{7.69}  & 3.3x   & 1.5x   & 1x       & 1.4x  & 1.9x    & \textcolor{black}{8.7x}     
                             
\end{tabular}
\label{tab:tpch_speedup}
\end{table}
\begin{table}[h]
\caption{The average runtimes of selected TCP-H queries with global and global scalar aggregation on SF-75 shown in seconds (fastest is highlighted in bold). }
\begin{tabular}{l|lllllll}
\toprule
SF-75 & TAG\_tg & psql & \dboSmall & \dboSmall\_im & \dbmd & \dbmd\_non & spark\_sql \\
\midrule
\texttt{q1}       & 32.1  & 54.6	& 13.3	& \textbf{7.6}	& 10.4	& 11.2  &  \textcolor{black}{16.6}     \\
\texttt{q6}       & 1.8  & 10.5	 & 1.6	& \textbf{0.6}	& 4.4	& 4.7    &  \textcolor{black}{1.7} \\
\texttt{q7}       & 6.8  & 24.8	& 5.6	& \textbf{3.5}	& 15.5	 & 4.8    & \textcolor{black}{40.5}  \\
\texttt{q9}       & 13.3  & 28.6	& 12.4	& 10.6	& \textbf{7.8}	& 9.2    & \textcolor{black}{50.2}  \\
\texttt{q16}       & 2.8  & 47.4	& 1.6	& \textbf{1.2}	& 2.4	 &2.4   & \textcolor{black}{58.6}  \\
\texttt{q19}       & 0.7  & 1.5	& \textbf{0.5}	& 1.1	& 3.2	& 3.8  &  \textcolor{black}{9.1}
                             
\end{tabular}
\label{tab:tpch_ga}
\end{table}
\paragraph{GA and Scalar GA Queries}
TAG-join does relatively worse on queries with global aggregation (GA), where $GROUP$ $BY$ consists of  multiple attributes that are not functionally dependent. Global aggregation requires a global data structure to which all the active vertices send their values
for aggregation (these are called {\em global aggregators} in the BSP model,
and are offered in TigerGraph). Global aggregators introduce a bottleneck, 
affecting query performance. Table~\ref{tab:tpch_ga} lists runtimes of selected queries (q1,q7,q9 and q16) with such GA, where we can observe that TAG-join is only consistently faster than PostgreSQL by 1.6-16x.  However, when filtering conditions of a query are very selective, i.e. small number of vertices need to write to the global structure, then TAG-join is quite competitive with \dbm as in the example of query q16. 

Our approach performs worse on query q1 compared to \dbo, \dbm and \revise{Spark SQL} (slower by 2.4-4x), however it is faster than PostgreSQL by 1.7x. This query is not a join query, but a single table scan where multiple scalar aggregations are computed over multiple columns.
On queries that compute scalar aggregates, where a single value is produced, TAG-join does well by outperforming most of the relational systems. Examples of scalar aggregation are q19 and q6 in Table~\ref{tab:tpch_ga}. Note that query q6 involves a scan of a single table in order to compute the scalar aggregate, \dbo in-memory is the fastest among the systems. This improvement in performance again can be explained by optimizations on data scans and aggregations that are enabled by compressed columnar format. The next fastest are TAG-join, \dbo (row store) \revise{and Spark SQL}, showing comparable performances.

\begin{table}[t]
\caption{Number of TPC-DS queries where TAG-join approach outperforms, shows competitive or worse performance against each of the relational systems at SF-75. Total number of queries is 84.
}

\begin{tabular}{l|lll}
\toprule
\#queries  & outperforms & competitive & worse \\
\midrule
psql       & 84          & -           & -     \\
\dboSmall     & 74          & 4           & 4     \\
\dboSmallIM & 64          & 3           & 17    \\
\dbmd      & 53          & 22          & 9     \\
\dbmnon & 64          & 12          & 8    \\
\revise{spark\_sql} & \revise{73}          & \revise{5}          & \revise{6}
\end{tabular}
\label{tab:num_outperform}
\end{table}
\subsection{\textcolor{black}{Single-Server} TPC-DS Results}\label{sec:centralized-tpc-ds}
Figure \ref{fig:total_runtimes}(b) shows the aggregate run times of TPC-DS queries over a dataset of different scale factors.

\revise{\dbm and \dbo employ an optimization technique such as bitmap filtering, which is most effective on star and snowflake schemes (i.e. TPC-DS). Bitmap filters are created from dimension tables and pushed down to the fact tables, which helps to reduce the cost of query processing and bring dramatic gains in query performance as a result. This technique is widely used in commercial databases ~\cite{bitmap_1, bitmap_2, bitmap_3}.     
Spark SQL often uses a broadcast join for star/snowflake joins, when relatively small dimension tables are broadcasted to all executors.
}

TAG-join on TigerGraph (TAG\_tg) performs the best on TPC-DS queries, 
followed by \dbo in-memory column store (\dboSmallIM), \dbo row-store (\dboSmall),
\dbm with clustered PK (\dbmd) and without (\dbmnon), then 
\revise{Spark SQL} and PostgreSQL (psql).
TAG-join is consistently faster than PostgreSQL on all 84 queries. It demonstrates better (or competitive) performance on $95\%$ of queries against \dbo, on $80\%$ of queries against \dbo
in-memory column store, \revise{on $93\%$ against Spark SQL } and on $90\%$ against \dbm (both clustered and non-clustered primary key settings). On the more detailed breakdown on Table~\ref{tab:num_outperform},we can observe that TAG-join outperforms relational systems on the majority of the TPC-DS queries. Runtimes of all individual queries for all scale factors are shown in Tables \ref{tab:tpcds_75},\ref{tab:tpcds_50},\ref{tab:tpcds_30} below.
  \begin{figure*}[t]
  \centering
  \includegraphics[width=\linewidth]{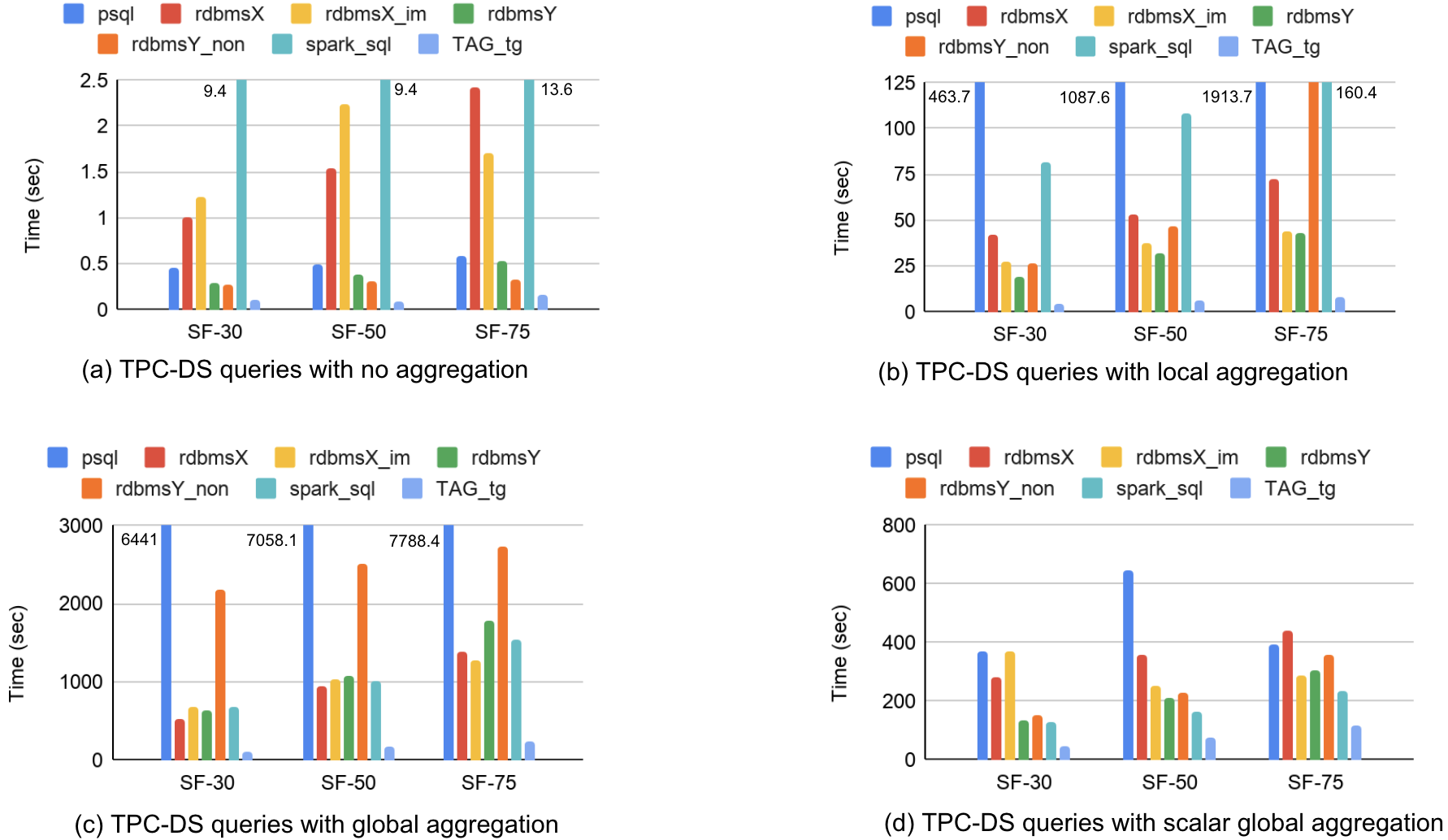}
  \caption{The aggregate runtimes of TPC-DS queries broken down into groups based on the aggregation type: no aggregation, local, global and scalar global. On subfigures (b) and (c) PostgreSQL and \textcolor{black}{Spark SQL} runtimes do not fit to show on the same scale with the rest, so the numbers are shown next to the bars.}
  \Description{}
  \label{fig:tpcds_agg}
\end{figure*}
Most of the TPC-DS queries involve aggregation, including local, global and scalar. In Figure~\ref{fig:tpcds_agg} we present aggregate times of queries across all systems by breaking down the queries into 4 groups based on the type of the aggregation.
Table~\ref{tab:tpcds_speedup} demonstrates the speedups of TAG-join on some individual queries.
\begin{table}[t]
\caption{The average runtimes of selected TCP-DS queries on SF-75 for TAG-join, shown in seconds, and it's relative speedups over relational engines. }
\begin{tabular}{l|cccccccc}
\toprule
SF-75 & TAG\_tg & psql & \dboSmall & \dboSmall\_im & \dbmd & \dbmd\_non & spark\_sql\\
\midrule

\textbf{No agg}    & \textbf{}       &        &        &            &       &       &     \\

\texttt{q37}       & \textbf{0.042}  & 3.6x   & 26.2x  & 18.1x      & 2.5x  & 2.3x  &  \textcolor{black}{95.9x}  \\
\texttt{q82}       & \textbf{0.04}   & 6.5x   & 27.1x  & 16.2       & 4.9x  & 2.8x  &  \textcolor{black}{164.5x}   \\
\texttt{q84}       & \textbf{0.075}  & 2.3x   & 3.2x   & 4.0x       & 2.9x  & 1.5x &  \textcolor{black}{39.9x}    \\
\midrule
\textbf{Local}        & \textbf{}       &        &        &            &       &     &       \\
\texttt{q7}        & \textbf{0.747}  & 9.9x   & 17.5x  & 1.7x       & 2.7x  & 3.04x  & \textcolor{black}{26.8x}   \\
\texttt{q12}       & \textbf{0.21}   & 12.6x  & 4.6x   & 3.04x      & 7.5x  & 6.6x  &  \textcolor{black}{12.6x}   \\
\texttt{q15}       & \textbf{0.72}   & 7.2x   & 7.3x   & 7.1x       & 1.1x  & 1.2x  &  \textcolor{black}{25.8x}   \\
\texttt{q20}       & \textbf{0.275}  & 9.6x   & 4.3x   & 2.9x       & 7.5x  & 6.5x  & \textcolor{black}{8.5x}    \\
\texttt{q33}       & \textbf{0.83}   & 13.1x  & 4.5x   & 5.7x       & 2.9x  & 3.5x  & \textcolor{black}{6.8x}    \\
\texttt{q50}       & \textbf{0.205}  & 7.8x   & 11.4x  & 5.02x      & 48.6x & 510.6x &  \textcolor{black}{78.5x}  \\
\texttt{q56}       & \textbf{0.451}  & 30.8x  & 7.5x   & 11.4x      & 3.6x  & 15.1x &  \textcolor{black}{11.6x}   \\
\texttt{q58}       & \textbf{1.244}  & 6.7x   & 3.2x   & 2.7x       & 7.6x  & 4.4x   & \textcolor{black}{4.5x}   \\
\texttt{q60}       & \textbf{0.916}  & 45.1x  & 7.8x   & 8.1x       & 3.4x  & 3.7x   & \textcolor{black}{6.5x}   \\
\texttt{q98}       & \textbf{0.44}   & 14.8x  & 3.3x   & 3.4x       & 7.9x  & 5.4x   &  \textcolor{black}{8.5x}  \\
\midrule
\textbf{Global}        & \textbf{}       &        &        &            &       &     &       \\
\texttt{q22}       & \textbf{3.551}  & 15.2x  & 1.4x   & 2.1x       & 1x    & 1x     & \textcolor{black}{1.9x}   \\
\texttt{q32}       & \textbf{0.047}  & 174.2x & 31.8x  & 5.1x       & 1.3x  & 2.3x   &  \textcolor{black}{53.3x}  \\
\texttt{q45}       & \textbf{0.234}  & 17.02x & 63.7x  & 83.6x      & 9.4x  & 9.7x   &  \textcolor{black}{51.9x}  \\
\texttt{q69}       & \textbf{1.317}  & 8.1x   & 4.6x   & 6.1x       & 2.1x  & 2.8x  &  \textcolor{black}{6.9x}   \\
\texttt{q74}       & \textbf{5.877}  & 13.1x  & 6.5x   & 2.9x       & 1.6x  & 1.5x  &  \textcolor{black}{7.1x}   \\
\texttt{q94}       & \textbf{0.185}  & 5.4x   & 7.02x  & 5.1x       & 1.6x  & 2.3x   &  \textcolor{black}{50.3x}                         
\end{tabular}
\label{tab:tpcds_speedup}
\end{table}
\paragraph{Queries without Aggregation}
As shown in Table~\ref{tab:tpcds_speedup}, TAG-join achieves 1.5-27x speedup over relational databases \revise{and 40-164x speedup over Spark SQL} on individual queries that do not use any aggregation function, i.e. select-project-join queries. There are only 3 queries of this type. 

\paragraph{LA Queries}
TAG-join does very well on queries involving local aggregation, as observed in the TPC-H results and confirmed on the 15 TPC-DS queries whose aggregation is local. 
TAG-join consistently outperforms PostgreSQL with an aggregate speedup of two orders of magnitude, and commercial relational systems with an aggregate speedup of 5-16x. We can observe the speedups of TAG-join on selected individual queries in Table~\ref{tab:tpcds_speedup}.
The highest speedups on individual LA queries are observed on queries that use a 
WITH clause in order to union results of subquery blocks, where each block joins at least 3 dimensions with a different fact table (e.g. q56, q33, q60). 
On these queries TAG-join is faster by 5-45x than PostgreSQL, by 3-8x than \dbo, by 3-11x than \dbo with in-memory column store, by 3-7.6x than \dbm with clustered primary key, by 3-15x than \dbm with non-clustered primary key, \revise{by 6.5-11x than Spark SQL}.
Queries q98, q20 and q12 are examples of queries with a PARTITION BY clause on one attribute (see Table~\ref{tab:tpcds_speedup}).

\paragraph{GA and Scalar GA Queries }
TAG-join does very well on most of the queries with global aggregation, showing 5-30x speedup in aggregate. There are 66 queries (out of 84) with GA or scalar GA. Most of these queries have quite selective filter conditions, leaving a smaller number of vertices active (relatively to the total number of vertices) that need to aggregate their values into the same global structure, and thus allowing to achieve quite competitive performance with relational systems. Examples of such queries with GA are in Table~\ref{tab:tpcds_speedup} (see queries q22, q45, q69 and q74). Note that these queries include roll-up aggregations (e.g. q18) to compute subtotal aggregate values of each group by key. This functionality is not offered 
out-of-the-box by the TigerGraph engine, but it can be simulated using multiple global structures with different keys.  

TAG-join shows a good performance on queries with global scalar aggregation achieving 2-3x performance improvement in aggregate over most of the relational engines, with exception of \dbo in-memory, which shows to be quite efficient in computing scalar aggregations over single column values. Queries q32 and q94 in Table~\ref{tab:tpcds_speedup} are examples of scalar global aggregation.

There are 19 out of total 84 TPC-DS queries, with either global or global scalar aggregation, where TAG-join performs poorly compared to \revise{relational engines, including \dbo, \dbm and Spark SQL} ( see full results in Tables \ref{tab:tpcds_75},\ref{tab:tpcds_50},\ref{tab:tpcds_30} below). On these queries we can observe a 2-12x speedup 
over TAG-join. TAG-join loses the most to \dbo with in-memory column store layout, see Table~\ref{tab:num_outperform}. \textbf{\dboSmallIM} enables an optimization called 'in-memory aggregation' which is especially beneficial on star queries, where multiple smaller dimensions are joined with a large fact table.

\subsection{Single-Server Memory Usage Results}\label{sec:mem_usage}
\subsubsection{Methodology}
We measured memory usage during workload execution with warm caches at a one second interval, and then reported a peak usage.
We read information from $/proc$ file system, which stores information about all processes currently running, including their memory usage. 

With automatic memory management enabled, \dbo allocates shared memory area through in-memory file system. The shared memory area includes buffer pool and in-memory column store.
PostgreSQL's buffer pool ($shared\_buffers$) is part of shared memory area as well.
Thus, in order to capture the full memory usage results for \dbo and PostgreSQL we need take into account the amount of shared memory (i.e. buffer pool) that is used during query execution.


We used $smem$ tool, that essentially pulls information from $/proc/\$\$/smaps$, which provides more detailed memory usage information.

\begin{table}[t]
\caption{\revise{Peak RAM usage of all systems during workload execution at SF-75.}}
\begin{tabular}{l|cccccc}
\toprule
       & psql    & \dboSmall  & \dboSmallIM & \dbmd   & spark\_sql & TAG\_tg \\
\midrule
TPC-H  &  65.9 \%    & 57.1 \%    & 51.2 \%       & 55.1 \%    & 57.4 \%    & 53.8 \% \\
TPC-DS & 61.7 \% & 49.8 \% & 43.5 \%    & 54.3 \% & 68.1 \%   & 52.9 \%
\end{tabular}
\label{tab:ram_usage}
\end{table}
\subsubsection{Results}
Results are summarized in Table~\ref{tab:ram_usage} for TPC-H and TPC-DS queries.
For \dbm the numbers are shown only for clustered PK,
being the same for non-clustered PK. 
We only show results for SF-75, but the numbers are proportional for SF-30 and SF-50.
Notice that TAG\_tg' memory performance is similar to \dbm and \dbo row store.
\dbo IM does better, but not by a game-changing margin:
9.4\% on TPC-DS (where TAG\_tg is faster though) and a negligible 2.6\% on TPC-H.

\subsection{Distributed Experiments}\label{sec:distr_exp}
In a cluster setting, we compared TAG-join implementation on TigerGraph 3.1 against Spark SQL/Spark 3.0.1. Experiment results are summarized in Figure~\ref{fig:distr_summary} in terms of aggregate runtimes of queries and total network traffic.
\subsubsection{Experiment Setup}
We run experiments on an Amazon EC2 cluster of 6 machines, each machine has 2.50GHz Intel Xeon Platinum 8259CL processor with 8 cores and 2 threads per core (i.e. 16 vCPU count), 64 GB of memory and 200GB SSD drive. All machines are running Ubuntu 18.04.

We used TigerGraph's default automatic partitioning of the input among the machines. We did not try to optimize or tune partitioning in this set of experiments, since we got good performance as is.
In TigerGraph we operate in a distributed query mode, in which graph traversals are executed in parallel on all machines, and the output is then gathered at one machine, where the given query is started. Otherwise, in a default mode TigerGraph selects one machine to execute a query, and data from other machines is copied to it for processing.   No additional tuning is done for TigerGraph.

A Spark application consists of a driver process and a number of executor processes distributed across machines in a cluster. Executor processes are launched by the driver process, read data from distributed files systems (S3 bucket in our experiments) and execute given query. Each executor process can run multiple tasks in parallel.
For Spark we tuned executor configurations like number of executors, executors cores and  executors memory.
We found 3 executors per machine, where each executor is assigned 5 cores and 16GB of memory gave the best performance for Spark in our experiments.
We use Parquet \cite{parquet}, a compressed columnar file format, as a data source, for which Spark supports column pruning and pushing down filter predicates. Spark SQL also offers an in memory caching of a data, which we use in our experiments as well. 
\subsubsection{Datasets and Queries}
We use TPC-H and TPC-DS benchmarks, as in a single machine setting, at scale factor 75. We run the same set of queries, i.e. 22 TPC-H queries and 84 TPC-DS queries. Each query is run three times, and we report the average runtime.
\begin{figure}[t]
  \centering
  \includegraphics[width=0.7\linewidth]{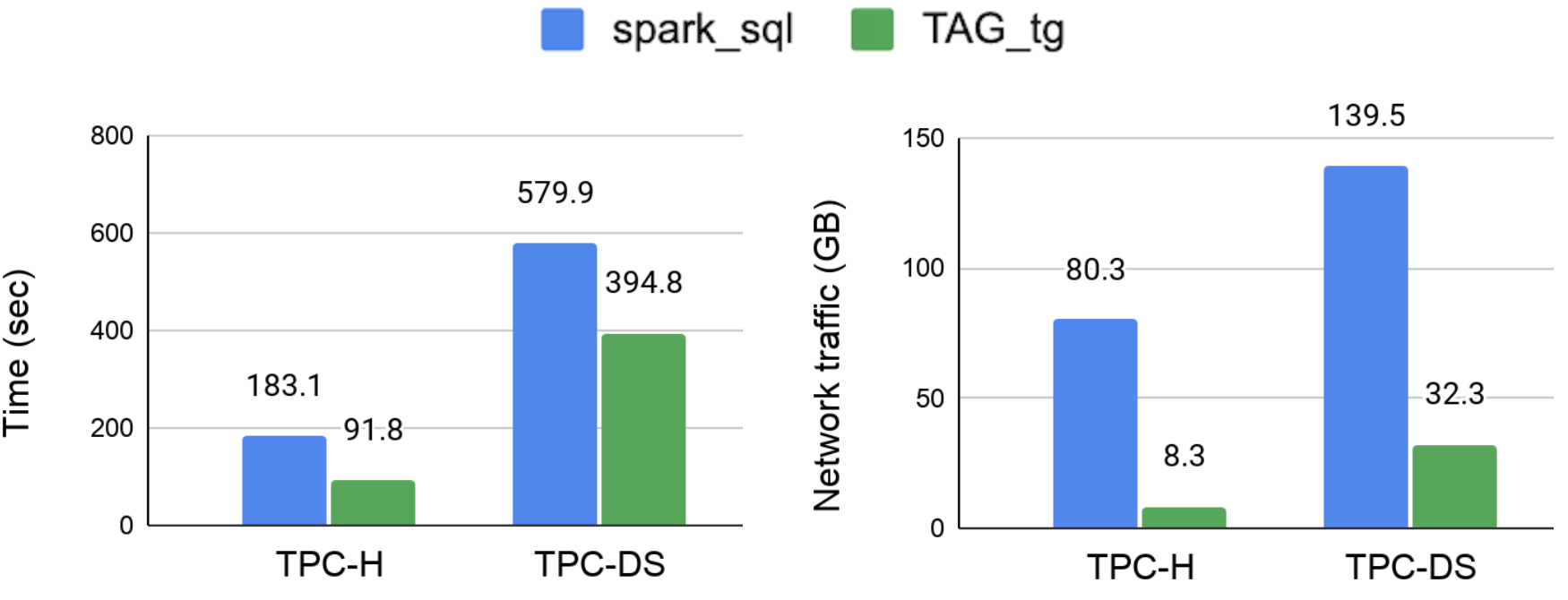}
  \caption{\revise{Summary of distributed experiments: aggregate time and network traffic.}}
  \label{fig:distr_summary}
  \Description{}
\end{figure}
\subsubsection{Results}
Figure~\ref{fig:distr_summary} shows the aggregate runtimes of TPC-H and TPC-DS queries. On TPC-H queries TAG-join in aggregate 2x faster than Spark SQL, and 1.46x faster on TPC-DS queries. Similarly to the centralized setting (single machine) experiments, the best performance is observed on queries without aggregation, with local aggregation and correlated subqueries.  

On TPC-H queries, TAG-join is faster than Spark SQL on 17 queries and competitive on 3 queries (out of total 22 queries). For example, on LA queries such as q3, q4, q5 and q10 the speedup ranges from 1.6x to 4x. The biggest speedup of 17x over Spark SQL is observed on q17, which contains a correlated a subquery.
Most queries with GA or scalar GA perform well using TAG-join 
except for q6 and q13 where Spark SQL is faster by 1.4-2.5x. Individual runtimes of all TPC-H queries are shown in Table~\ref{tab:dist_tpch}.

Out of 84 TPC-DS queries, TAG-join is either competitive or outperforms Spark SQL on 64 queries.
On queries without aggregation the speedup is 3.4-5.5x, while on queries with LA TAG-join achieves up to 7.6x speedup. 
TAG-join performs well on most of the queries with either GA or single GA.
Spark SQL is only faster on 20 queries, where either GA or single GA is computed. We observed the same in a single machine setting. In order to compute the final result all active vertices need to write into a single global accumulator, and with a lot of active vertices, this can significantly degrade the performance, since no parallelism is gained.
Individual runtimes of all TPC-DS queries are shown in Table~\ref{tab:dist_tpcds}.

We track network usage during query execution on each machine in the cluster using $sar$ tool, and record the total number of bytes received and transmitted during execution of all queries for each benchmark.
Figure~\ref{fig:distr_summary} shows the total incoming traffic, i.e. incoming traffic summed over all machines in the cluster. We only report incoming traffic as it coincides with the total outgoing traffic.
Spark SQL incurs 9x more traffic on TPC-H benchmark and 4x more traffic on TPC-DS benchmark. Spark uses broadcast join or shuffle join, which requires replication of data over many partitions, thus more network traffic.

\section{Conclusion and Future Work}\label{sec:conclusion}
We have shown that the TAG encoding and our TAG-join algorithm
combine to unlock the potential of vertex-centric SQL evaluation \revise{to exploit
both intra- and inter-machine parallelism}.
By running full TPC SQL queries we have proven that our vertex-centric approach is
compatible with executing RA operations beyond joins.
The observed performance constitutes very promising evidence for the relevance of vertex-centric approaches to SQL evaluation.

\revise{
From the SQL user’s perspective, our experiments show that in single-server data warehousing settings, vertex-centric evaluation can clearly outperform even leading
commercial engines like \dbo IM. In TPC-H workloads, comparison to \dbo IM
depends on the kind of aggregation performed, while our approach is competitive with
or superior to the other relational engines. In a distributed cluster, 
our TAG-join implementation outperforms Spark SQL on both TPC benchmarks.
}

Our main focus has been on join evaluation and we have only scratched the surface of inter-operator optimizations,
\revise{confining ourselves to those inspired by the relational setting 
(like pushing selection, projection and aggregations before joins). 
We plan to explore optimizations specific
to the vertex-centric model.}

\revise{If the value domain is continuous and the database is constantly being updated, 
the TAG encoding would prescribe creating a new attribute vertex for virtually
each incoming value, which is impractical. Applying our vertex-centric paradigm to this
scenario is an open problem which constitutes an appealing avenue for future work.
}

\bibliographystyle{ACM-Reference-Format}
\bibliography{sample-sigconf}

\clearpage
\appendix
\section{Full Experimental Results}\label{sec:appendix_results}

\begin{table*}[h]
\caption{Average runtimes of TPC-H queries for SF-75, shown in seconds.}
\begin{tabular}{c|c|c|c|c|c|c|c}
 \toprule
SF-75    & psql    & \dboSmall & \dboSmallIM & \dbmd  & \dbmnon & spark\_sql & TAG\_tg     \\
\midrule
q1    & 54.609  & 13.305                   & 7.678                      & 10.405               & 11.266                 & 16.613                                     & 32.129  \\
q2    & 24.588  & 1.876                    & 0.615                      & 1.09                 & 1.069                  & 18.589                                     & 0.645   \\
q3    & 19.162  & 4.493                    & 2.576                      & 7.465                & 7.756                  & 21.565                                     & 4.288   \\
q4    & 8.235   & 3.69                     & 1.252                      & 4.167                & 4.887                  & 15.144                                     & 1.72    \\
q5    & 50.913  & 7.463                    & 8.101                      & 3.016                & 5.477                  & 36.516                                     & 4.765   \\
q6    & 10.551  & 1.596                    & 0.582                      & 4.472                & 4.749                  & 1.789                                      & 1.837   \\
q7    & 24.849  & 5.626                    & 3.57                       & 15.52                & 4.832                  & 40.594                                     & 6.889   \\
\end{tabular}
\label{tab:tpch_75}
\end{table*}

\begin{table*}[]
\caption{Average runtimes of TPC-H queries for SF-50, shown in seconds.}
\begin{tabular}{c|c|c|c|c|c|c|c}
 \toprule
SF-50    & psql    & \dboSmall & \dboSmallIM & \dbmd  & \dbmnon & spark\_sql & TAG\_tg     \\
    \midrule
q1    & 40.534  & 9.258                    & 5.239                      & 7.533                & 7.546                  & 7.824                 & 21.499  \\
q2    & 16.149  & 1.718                    & 1.17                       & 1.077                & 1.191                  & 18.408                & 0.429   \\
q3    & 13.688  & 3.043                    & 1.94                       & 5.853                & 6.335                  & 16.309                & 3.05    \\
q4    & 5.244   & 2.912                    & 0.928                      & 3.02                 & 3.168                  & 10.380                & 1.183   \\
q5    & 31.802  & 5.759                    & 5.9                        & 1.878                & 3.422                  & 26.263                & 3.323   \\
q6    & 7.619   & 1.251                    & 0.441                      & 2.828                & 3.111                  & 1.415                 & 1.27    \\
q7    & 15.88   & 3.495                    & 2.354                      & 2.406                & 6.294                  & 28.173                & 4.896   \\
\end{tabular}
\label{tab:tpch_50}
\end{table*}

\begin{table*}[t]
\caption{Average runtimes of TPC-H queries for SF-30, shown in seconds.}
\begin{tabular}{c|c|c|c|c|c|c|c}
 \toprule
 SF-30 & psql    & \dboSmall & \dboSmallIM & \dbmd  & \dbmnon & spark\_sql & TAG\_tg     \\
 \midrule
q1                        & 24.324 & 7.196                    & 3.191                      & 4.667                & 4.55                   & 6.322                & 12.762  \\
q2                        & 9.531  & 1.396                    & 0.69                       & 0.772                & 0.878                  & 10.476               & 0.263   \\
q3                        & 7.758  & 2.019                    & 1.124                      & 5.826                & 5.566                  & 11.254               & 1.882   \\
q4                        & 3.194  & 2.664                    & 0.653                      & 1.632                & 1.923                  & 7.200                & 0.779   \\
q5                        & 23.511 & 7.203                    & 3.918                      & 1.284                & 2.101                  & 14.015               & 2.035   \\
q6                        & 4.496  & 1.237                    & 0.331                      & 1.921                & 1.939                  & 0.884                & 0.768   \\
q7                        & 10.533 & 2.86                     & 1.638                      & 5.982                & 1.954                  & 14.463               & 2.985   \\
\end{tabular}
 \label{tab:tpch_30}
\end{table*}

\begin{table*}[t]
\caption{Average runtimes of TPC-DS queries for SF-75, shown in seconds. "-" indicates timeout.}
\begin{tabular}{c|c|c|c|c|c|c|c}
 \toprule
 SF-75 & psql    & \dboSmall & \dboSmallIM & \dbmd  & \dbmnon & spark\_sql & TAG\_tg     \\
 \midrule
\textbf{No agg} &        &        &            &       &            &        &  \\
q37                              & 0.152   & 1.099                    & 0.76                       & 0.104                & 0.097                  & 4.030                & 0.042   \\
q82                              & 0.262   & 1.083                    & 0.648                      & 0.198                & 0.111                  & 6.583                & 0.04    \\
q84                              & 0.172   & 0.238                    & 0.302                      & 0.221                & 0.112                  & 2.996                & 0.075   \\

\midrule 
\textbf{Local}  &        &        &            &       &            &        &   \\
q12                              & 2.655   & 0.982                    & 0.64                       & 1.59                 & 1.394                  & 2.660                & 0.21    \\
q15                              & 5.185   & 5.269                    & 5.098                      & 0.793                & 0.896                  & 18.637               & 0.72    \\
\end{tabular}
 \label{tab:tpcds_75}
\end{table*}

\begin{table*}[t]
\begin{tabular}{c|c|c|c|c|c|c|c}
 \toprule
 SF-75  & psql    & \dboSmall & \dboSmallIM & \dbmd  & \dbmnon & spark\_sql  & TAG\_tg     \\
 \midrule
\textbf{Global} &         &         &            &          &            &       &   \\
q43                              & 8.95    & 20.518                   & 1.321                      & 1.991                & 2.059                  & 2.399                & 3.059   \\
q45                              & 3.983   & 14.918                   & 19.577                     & 2.202                & 2.29                   & 12.164               & 0.234   \\
q46                              & 8.389   & 9.043                    & 1.696                      & 3.375                & 3.292                  & 11.247               & 3.025   \\
q48                              & 12.489  & 15.459                   & 1.12                       & 7.426                & 1.662                  & 17.199               & 2.013   \\
q5                               & 13.979  & 5.446                    & 54.648                     & 2.601                & 3.412                  & 12.415               & 1.799   \\
q52                              & 7.519   & 5.64                     & 4.204                      & 0.204                & 0.619                  & 2.687                & 0.235  \\
\end{tabular}
 \label{tab:tpcds_75_1}
\end{table*}

\begin{table*}[t]
\caption{Average runtimes of TPC-DS queries for SF-50, shown in seconds. "-" indicates timeout.}
\begin{tabular}{c|c|c|c|c|c|c|c}
 \toprule
 SF-50 & psql    & \dboSmall & \dboSmallIM & \dbmd  & \dbmnon & spark\_sql & TAG\_tg     \\
 \midrule
 \textbf{No agg} &        &        &            &       &            &         & \\
q37                              & 0.14     & 0.114                    & 1.813                      & 0.125                & 0.088                  & 2.769                & 0.022   \\
q82                              & 0.223    & 0.11                     & 0.138                      & 0.154                & 0.1                    & 4.304                & 0.024   \\
q84                              & 0.133    & 1.311                    & 0.285                      & 0.105                & 0.12                   & 2.370                & 0.038   \\
\midrule
\textbf{Local} &        &        &            &       &            &         \\
q12                              & 1.922    & 0.651                    & 0.499                      & 1.367                & 1.294                  & 2.573                & 0.109   \\
q15                              & 3.539    & 3.396                    & 2.912                      & 0.522                & 0.704                  & 6.799                & 0.467   \\
\end{tabular}
 \label{tab:tpcds_50}
\end{table*}

\begin{table*}[t]
\begin{tabular}{c|c|c|c|c|c|c|c}
 \toprule
 SF-50 & psql    & \dboSmall & \dboSmallIM & \dbmd  & \dbmnon & spark\_sql & TAG\_tg     \\
 \midrule
\textbf{Global} &        &        &            &       &            &         \\
q43                              & 5.608    & 13.087                   & 1.012                      & 1.57                 & 1.251                  & 1.754                & 1.657   \\
q45                              & 2.766    & 9.558                    & 8.194                      & 2.014                & 2.175                  & 4.742                & 0.157   \\
q46                              & 6.878    & 6.122                    & 1.601                      & 3.107                & 2.787                  & 7.070                & 2.749   \\
q48                              & 8.718    & 10.597                   & 0.859                      & 4.791                & 1.146                  & 11.115               & 1.453   \\
q5                               & 10.071   & 3.581                    & 49.81                      & 1.954                & 2.382                  & 9.743                & 1.239   \\
q52                              & 5.196    & 3.717                    & 9.264                      & 0.161                & 0.46                   & 1.521                & 0.148  \\
\end{tabular}
 \label{tab:tpcds_50_1}
\end{table*}

\begin{table*}[t]
\caption{Average runtimes of TPC-DS queries for SF-30, shown in seconds. "-" indicates timeout.}
\begin{tabular}{c|c|c|c|c|c|c|c}
 \toprule
 SF-30 & psql    & \dboSmall & \dboSmallIM & \dbmd  & \dbmnon & spark\_sql & TAG\_tg     \\
 \midrule
\textbf{No agg} &        &        &            &       &            &         \\
q37                              & 0.139   & 0.435                    & 0.519                      & 0.064                & 0.084                  & 2.927                & 0.031   \\
q82                              & 0.218   & 0.437                    & 0.627                      & 0.127                & 0.095                  & 3.805                & 0.03    \\
q84                              & 0.107   & 0.135                    & 0.087                      & 0.097                & 0.098                  & 2.698                & 0.043   \\
\midrule
\textbf{Local} &        &        &            &       &            &         \\
q12                              & 1.332   & 0.415                    & 0.366                      & 0.831                & 1.072                  & 2.302                & 0.101   \\
q15                              & 1.181   & 2.134                    & 3.284                      & 0.333                & 0.397                  & 2.544                & 0.358   \\
\end{tabular}
 \label{tab:tpcds_30}
\end{table*}

\begin{table*}[t]
\begin{tabular}{c|c|c|c|c|c|c|c}
 \toprule
 SF-30 & psql    & \dboSmall & \dboSmallIM & \dbmd  & \dbmnon & spark\_sql & TAG\_tg     \\
 \midrule
\textbf{Global} &        &        &            &       &            &         \\
q43                              & 4.358   & 16.057                   & 0.64                       & 0.816                & 0.844                  & 1.536                & 1.336   \\
q45                              & 2.043   & 5.561                    & 6.599                      & 1.927                & 2.05                   & 1.931                & 0.135   \\
q46                              & 5.262   & 3.474                    & 1.335                      & 2.941                & 2.461                  & 6.147                & 2.162   \\
q48                              & 6.132   & 11.558                   & 0.641                      & 2.751                & 0.72                   & 8.162                & 0.919   \\
q5                               & 6.753   & 2.35                     & 5.632                      & 1.328                & 1.595                  & 8.133                & 0.788   \\
q52                              & 3.589   & 2.442                    & 1.819                      & 0.127                & 0.343                  & 1.219                & 0.129   \\
\end{tabular}
 \label{tab:tpcds_30_1}
\end{table*}

\begin{table*}[t]
\caption{Aggregate time in seconds}
\begin{tabular}{l|ccc}
\toprule
           & \multicolumn{3}{c}{TPC-H (TPC-DS)}                                                        \\
           & \multicolumn{1}{c}{SF-30} & \multicolumn{1}{c}{SF-50} & \multicolumn{1}{c}{SF-75} \\
\midrule
psql      & 358.9 (7275.7) &592.6 (8789.3)  &796.2 (10094.3)\\
\dboSmall     & 78.6 (842.6)  & 88.8 (1357.7)  & 104.1 (1888.6)\\
\dboSmallIM  & 32.5 (1065.7)  & 52.2 (1317.1)   & 74.5 (1611.2)\\
\dbmd      & 59.2 (787.1)   & 80.4 (1328.2)    & 132.7 (2117.7)\\
\dbmnon & 70.7 (2349.6)  & 87.6 (2780.8)    & 139.8 (3230.9)\\
\revise{spark\_sql}     & \revise{248.9 (906.4)}   & \revise{391.5 (1282.2)}  & \revise{572.9 (1955.2)}\\
TAG\_tg     & 53.4 (149.1)   & 85.1 (246.3)  & 121.3 (357.3)
\end{tabular}
\label{tab:total_runtimes}
\end{table*}

\begin{table*}[h]
\caption{Size of the \dbo in-memory column store segments, as well as original data size, shown in GB.}
\begin{tabular}{ll|rr}
\toprule
       &       & \multicolumn{1}{l}{data size} & \multicolumn{1}{l}{in-memory size} \\
\midrule
 TPC-H      & SF-30 & 36                            & 24                                 \\
       & SF-50 & 59                            & 42                                 \\
       & SF-75 & 89                            & 63                                 \\
\midrule
TPC-DS       & SF-30 & 23                            & 16                                 \\
       & SF-50 & 38                            & 26                                 \\
       & SF-75 & 57                            & 40                                
\end{tabular}
\label{tab:load_im}
\end{table*}

\begin{table*}[h]
\caption{Average runtimes of TPC-H queries for SF-75 in a cluster setting, shown in seconds.}
\begin{tabular}{c|c|c}
 \toprule
 SF-75 & spark\_sql  & TAG\_tg \\
 \midrule
 \texttt{q1}                        & 4.79                           & 4.97                        \\
\texttt{q2}                        & 8.76                           & 3.44                        \\
\texttt{q3}                        & 6.91                           & 4.12                        \\
\texttt{q4}                        & 4.96                           & 2.09                        \\
\texttt{q6}                        & 0.88                           & 2.23                        \\
\texttt{q8}                        & 11.16                          & 4.21                        \\
\texttt{q9}                        & 14.90                          & 10.17                       \\
\end{tabular}
\label{tab:dist_tpch}
\end{table*}
\begin{table*}[h]
\caption{Average runtimes of TPC-DS queries for SF-75 in a cluster setting, shown in seconds.}
\begin{tabular}{c|c|c}
 \toprule
 SF-75 & spark\_sql  & TAG\_tg \\
 \midrule
\textbf{No agg} &        &     \\
\texttt{q37} & 1.345 & 0.387 \\
\texttt{q82} & 2.195 & 0.398 \\
\texttt{q84} & 1.038 & 0.722 \\
\midrule
\textbf{Local} &       &      \\
\texttt{q12} & 1.463 & 1.323 \\
\texttt{q15} & 4.970 & 2.333 \\
\texttt{q20} & 1.048 & 1.690 \\
\texttt{q26} & 3.262 & 1.395 \\
\texttt{q33} & 2.120 & 1.544 \\
\texttt{q50} & 5.370 & 0.972 \\
\texttt{q56} & 1.738 & 1.769 \\
\texttt{q58} & 3.074 & 3.616 \\
\texttt{q6}  & 4.449 & 2.285 \\
\texttt{q60} & 1.659 & 2.006 \\
\texttt{q7}  & 5.219 & 1.380 \\
\texttt{q83} & 1.612 & 0.986 \\
\texttt{q85} & 4.314 & 2.216 \\
\texttt{q93} & 8.760 & 1.154 \\
\texttt{q98} & 1.304 & 1.762 \\
\midrule
\textbf{Global} &       &      \\
\texttt{q1}  & 2.144   & 2.324  \\
\texttt{q10} & 9.444   & 3.693  \\
\texttt{q11} & 15.344  & 6.731  \\
\texttt{q13} & 6.917   & 2.164  \\
\texttt{q16} & 5.496   & 1.671  \\
\texttt{q18} & 6.665   & 4.745  \\
\texttt{q19} & 2.563   & 3.448  \\
\texttt{q2}  & 1.543   & 5.338  \\
\texttt{q21} & 1.020   & 1.006  \\
\texttt{q22} & 3.930   & 3.086  \\
\texttt{q25} & 10.814  & 11.810 \\
\texttt{q27} & 4.987   & 1.249  \\
\texttt{q28} & 4.097   & 1.332  \\
\texttt{q29} & 10.355  & 8.477 \\
\texttt{q3}  & 0.660   & 0.756  \\
\texttt{q30} & 2.531   & 5.327  \\
\texttt{q31} & 6.211   & 5.147  \\
\texttt{q32} & 1.032   & 0.892  \\
\texttt{q34} & 1.331   & 1.783  \\
\texttt{q35} & 5.571   & 7.067  \\
\texttt{q38} & 6.467   & 13.733 \\
\texttt{q4}  & 16.327  & 7.832  \\
\texttt{q40} & 3.579   & 2.184  \\
\texttt{q41} & 0.461   & 0.209  \\
\texttt{q42} & 0.689   & 1.160 
\end{tabular}
\label{tab:dist_tpcds}
\end{table*}

\begin{table*}[h]
\begin{tabular}{c|c|c}
 \toprule
 SF-75 & spark\_sql  & TAG\_tg \\
 \midrule
\textbf{Global} &       &      \\
\texttt{q43} & 1.029   & 3.002  \\
\texttt{q45} & 3.431   & 2.428  \\
\texttt{q46} & 2.478   & 5.212  \\
\texttt{q48} & 5.186   & 4.398  \\
\texttt{q5}  & 4.708   & 4.771  \\
\texttt{q52} & 0.906   & 1.065  \\
\texttt{q53} & 1.398   & 1.598  \\
\texttt{q54} & 5.429   & 3.182  \\
\texttt{q55} & 0.623   & 1.162  \\
\texttt{q59} & 2.485   & 39.986 \\
\texttt{q61} & 2.779   & 1.376  \\
\texttt{q62} & 1.025   & 6.380  \\
\texttt{q63} & 1.030   & 1.516  \\
\texttt{q64} & 23.212  & 4.824  \\
\texttt{q65} & 4.138   & 12.880 \\
\texttt{q66} & 2.488   & 4.165  \\
\texttt{q68} & 2.418   & 3.114  \\
\texttt{q69} & 2.443   & 3.462  \\
\texttt{q71} & 1.592   & 2.220  \\
\texttt{q72} & 213.122 & 21.935 \\
\texttt{q73} & 1.142   & 1.219  \\
\texttt{q74} & 10.398  & 6.625  \\
\texttt{q75} & 8.366   & 6.936  \\
\texttt{q76} & 1.856   & 2.570  \\
\texttt{q77} & 15.470  & 4.181  \\
\texttt{q78} & 15.553  & 50.814 \\
\texttt{q79} & 2.428   & 4.466  \\
\texttt{q80} & 15.490 & 3.998  \\
\texttt{q81} & 3.103  & 9.445  \\
\texttt{q87} & 6.761  & 13.436 \\
\texttt{q88} & 3.543  & 1.956  \\
\texttt{q89} & 1.428  & 1.813  \\
\texttt{q9}  & 2.115  & 4.877  \\
\texttt{q90} & 0.520  & 0.527  \\
\texttt{q91} & 1.135  & 0.599  \\
\texttt{q92} & 1.175  & 0.893  \\
\texttt{q94} & 2.838  & 1.570  \\
\texttt{q95} & 14.256 & 1.768  \\
\texttt{q96} & 0.435  & 0.617  \\
\texttt{q97} & 3.821  & 14.909 \\
\texttt{q99} & 1.075  & 13.591
 \end{tabular}
\end{table*}
 
\end{document}